\documentclass[10pt,journal,cspaper,compsoc]{IEEEtran}

\usepackage{color}
\usepackage{enumerate}
\usepackage{multirow}
\usepackage{diagbox}
\usepackage{subcaption}
\usepackage{makecell}
\usepackage{listings}

\usepackage{geometry}
\usepackage{caption} 
\usepackage{nohyperref}
\usepackage{url}

\usepackage{amsmath}
\usepackage{pifont}
\newcommand{\cmark}{\text{\ding{51}}}
\newcommand{\xmark}{\text{\ding{55}}}

\usepackage[nocompress]{cite}

\usepackage[font=bf]{caption}
\DeclareCaptionType{copyrightbox}
\usepackage{subcaption}

\usepackage{diagbox}

\ifCLASSINFOpdf
   \usepackage[pdftex]{graphicx}
   \graphicspath{{../pdf/}{../jpeg/}}
   \DeclareGraphicsExtensions{.pdf,.jpeg,.png}
\else
   \usepackage[dvips]{graphicx}
   \graphicspath{{../eps/}}
   \DeclareGraphicsExtensions{.eps}
\fi

\usepackage{multirow}

\usepackage{amsmath}
\usepackage{algorithm}
\usepackage{algorithmic}
\usepackage{array}
\usepackage{comment}

\usepackage{amssymb}

\lstdefinestyle{baseC}{
  language=C,
  tabsize=1,
  breaklines=true,
  basicstyle=\scriptsize\ttfamily\color{black},
  showstringspaces=false,
  moredelim=**[is][\color{red}]{@}{@},
}

\lstdefinestyle{baseJava}{
  language=Java,
  tabsize=1,
  breaklines=true,
  basicstyle=\scriptsize\ttfamily\color{black},
  showstringspaces=false,
  moredelim=**[is][\color{red}]{@}{@},
}

\usepackage{amssymb}
\usepackage{pifont}

\usepackage{tikz}
\usetikzlibrary{arrows,decorations.pathmorphing,decorations.text,backgrounds,fit,positioning,shapes.symbols,chains,calc,shadings,patterns,shapes,shapes.geometric,automata}
\tikzset{mynode/.style={draw=black,ellipse,inner sep=1pt,font=\scriptsize,anchor=south}}
\input{lstStyle}

\newcommand{\red}[1]{{\color{black}#1}}
\newcommand{\blue}[1]{{\color{black}#1}}
\newcommand{\green}[1]{{\color{black}#1}}

\makeatletter
\newcommand\notsotiny{\@setfontsize\notsotiny\@vipt\@viipt}
\makeatother

\ifCLASSOPTIONcompsoc
  \usepackage[nocompress]{cite}
\else
  \usepackage{cite}
\fi
\ifCLASSINFOpdf
\else
\fi
\begin{document}
%
\title{TFix$^+$: Self-configuring Hybrid Timeout Bug Fixing for Cloud Systems}
%
%
%
%

\author{Jingzhu He,
        Ting Dai,~\IEEEmembership{Member,~IEEE,}
        and Xiaohui Gu,~\IEEEmembership{Senior~Member,~IEEE}
\IEEEcompsocitemizethanks{\IEEEcompsocthanksitem Jingzhu He is with ShanghaiTech University. \protect\\
E-mail: hejzh1@shanghaitech.edu.cn

\IEEEcompsocthanksitem Ting Dai is with IBM Research.\protect\\
E-mail: Ting.Dai@ibm.com

\IEEEcompsocthanksitem Xiaohui Gu is with North Carolina State University.\protect\\
E-mail: xgu@ncsu.edu

}
\thanks{}}

 \markboth{IEEE Transactions on Dependable and Secure Computing (TDSC)}%
 {}
\IEEEcompsoctitleabstractindextext{%
\begin{abstract}
\red{Timeout bugs can cause serious availability and performance issues which are often difficult to fix due to the lack of diagnostic information. Previous work proposed solutions for fixing specific type of timeout-related performance bugs. 
In this paper, we present TFix$^+$, a self-configuring timeout bug fixing framework for automatically correcting two major kinds of timeout bugs (i.e., misused timeout bugs and missing timeout bugs) with dynamic timeout value predictions. 
TFix$^+$ provides two new hybrid schemes for fixing misused and missing timeout bugs, respectively. 
TFix$^+$ further provides prediction-driven timeout variable configuration based on runtime function tracing. 
We have implemented a prototype of TFix$^+$ and conducted experiments on 16 real world timeout bugs. 
Our experimental results show that TFix$^+$ can effectively fix 15 out of tested 16 timeout bugs.}

\end{abstract}

}

\maketitle

\IEEEdisplaynotcompsoctitleabstractindextext

%
\IEEEpeerreviewmaketitle


\section{Introduction}


Timeout is commonly used to handle unexpected failures in complex distributed systems. For example, when a server $A$ sends a request to another server $B$, $A$ can use the timeout mechanism to avoid endless waiting in case $B$ fails to respond. 
\blue{Timeout bugs can severely impact system availability and performance, causing system hang and performance degradation~\cite{gunawi2014bugs,huang2015understanding,timeoutstudy}. 
Our previous bug study~\cite{timeoutstudy} shows that 78\% of real world timeout bugs are caused by missing timeout mechanisms or mis-using timeout schemes (e.g., setting a too small or a too large timeout value).}
For example, a misused timeout bug caused Amazon DynamoDB to experience a five-hour service outage in 2015~\cite{dynamoDBtimeout}. The root cause of this bug is an improper timeout value setting under unexpected workload increase. 
Timeout bugs are often difficult to fix because of the lack of diagnostic information and the correct timeout value often depends on the runtime execution environments and application workloads.

Previous work \cite{timeoutfix, he2020hangfix} has proposed bug fixing schemes for specific bug types (e.g., hang bugs, data corruption bugs) using static code analysis methods. However, an effective generic timeout bug fixing solution requires both runtime knowledge (e.g., network bandwidth for setting a proper RPC timeout variable) and root cause function localization \cite{timeoutfix, he2020hangfix}. In this paper, we propose a hybrid approach to fixing both misused and missing timeout bugs. We introduce prediction-driven timeout configuration scheme to realize self-configuring fully automated timeout bug fixing.

\begin{figure}[!t]
\centering
\begin{lstlisting}[escapechar=|, frame=single, style=baseJava, tabsize=1]
//yarn-default.xml                     Yarn-1630(v2.2.0) 
			@RED@+@RED@ <property>
			@RED@+@RED@     <name>@BG@poll.timeout@BG@</name>
			@RED@+@RED@     <value>@BG@1130@BG@</value>
			@RED@+@RED@ </property>

//YarnClientImpl.java              
			@RED@+@RED@ private String @BG@POLL_TIMEOUT_KEY@BG@ = "poll.timeout";
			@RED@+@RED@ private long @BG@timeout@BG@ = conf.getInt(
			@RED@+@RED@							  POLL_TIMEOUT_KEY); 

142  public ApplicationId submitApplication(...)
143       throws YarnException, IOException {
						...
			@RED@+@RED@		@BG@long st = System.currentTimeMillis();@BG@
153   while(true) {
154    state = getAppReport(appId).getYarnAppState();
155    if (!state.equals(YarnAppState.NEW) && 
									!state.equals(YarnApplicationState.NEW_SAVING))
156     break;
							...   //checking elapsed time
			@RED@+@RED@			@BG@long elapsed= System.currentTimeMillis() - st;@BG@
			@RED@+@RED@			@BG@if (timeout > 0 && elapsed >= timeout) {@BG@
			@RED@+@RED@			@BG@	throw new TimeoutException("Timed out."@BG@
			@RED@+@RED@			@BG@							+"Breaking infinite polling!"); }@BG@
170   }
						...
177  }
\end{lstlisting}
\vspace*{-1\baselineskip}
\caption{
An example of missing timeout bug Yarn-1630 (v2.2.0). 
The {\tt submitApplication} function keeps polling its state but gets {\tt NEW},
hanging in an infinite loop.
``{\color{red}+}'' shows the patch produced by TFix$^+$. 
}
\label{fig:Yarn-1630}
\end{figure}

\subsection{A Motivating Example}
\label{sec:motivating}

To better understand how real-world timeout bugs happen, and how they can affect cloud services, 
we use the Yarn-1630 (v2.2.0)~\footnote{We use ``system name-bug \#" to denote different bugs.} bug as one example shown by Figure~\ref{fig:Yarn-1630}. 
This bug is caused by missing timeout settings on asynchronous polling operations in YarnClientImpl.
When submitting an application, the Yarn client periodically polls the ResourceManager to get the application status (line \#154), \blue{until the application is submitted successfully (line \#155).}
\blue{With a slow submission process in the ResourceManager end, the Yarn client gets stuck, causing system hang indefinitely.}

Figure~\ref{fig:Yarn-1630} also shows the patch generated by TFix$^+$.
\blue{To break the Yarn client from the infinite waiting while still guaranteeing that the asynchronous communications between the Yarn client and the ResourceManager can succeed during normal runs, TFix$^+$ inserts a timeout mechanism in the loop body (added lines between line \#156 to \#170) with user configurable timeout variables (added lines before line \#142). The automatically generated path by TFix$^+$ saves users diagnosing and development effort to fix the bug.}
\blue{To further ease the users' testing work and better accommodate their runtime environments, TFix$^+$} provides dynamic timeout value predictions based on runtime monitoring data. In this example, we configure the timeout value as 1130 ms based on the runtime available resources and our workloads.

\subsection{Contribution}

In this paper, we present TFix$^+$, a self-configuring hybrid timeout bug fixing system.
\blue{Compared with its preliminary version TFix~\cite{timeoutfix}, which focuses only on fixing misused timeout bugs using static analysis methods,
TFix$^+$ provides self-configuring holistic fixing solutions for both misused and missing timeout bugs. TFix$^+$ proposes dynamic timeout value prediction scheme using both runtime tracing and static taint analysis.} 
When a timeout bug is detected by a runtime bug detection tool such as TScope~\cite{hetscope}, TFix$^+$ executes a {\em drill-down bug analysis} protocol to automatically narrow down the root cause of the detected bug and produce bug fixing patches with sound timeout value recommendations. 
Specifically, TFix$^+$ first determines whether the detected timeout bug is caused by \blue{misused timeout bugs (i.e., incorrectly used timeout variables)} or \blue{missing timeout bugs (i.e., lacking timeout mechanisms)}. 
\blue{For misused timeout bugs, TFix$^+$ identifies timeout affected functions from an application performance trace and then uses those functions to pinpoint misused timeout variables using taint analysis.}
\blue{For missing timeout bugs, TFix$^+$ identifies the root cause as either an infinite loop or a blocking function call via stack trace analysis, and inserts different timeout mechanisms with configurable timeout variables, respectively.}
\blue{To achieve high efficiency and obtain strong robustness,}
TFix$^+$ recommends a proper timeout value \blue{to adapt to the user runtime environments} based on the historical execution time of the pinpointed timeout affected functions during normal runs. 
TFix$^+$ validates the generated patch by checking whether the bug still occurs under the same workload and all the test suites are passed after adopting the patch.
Specifically, our paper makes the following contributions.

\begin{itemize}
\item We describe a holistic {\em drill-down bug analysis} framework which can automatically narrow down the root cause of a timeout bug and provide a complete bug fixing patch. 
\item We describe a hybrid scheme that combines dynamic application performance tracing and static taint analysis to find the timeout affected function and localize the misused timeout variable when the bug is caused by mis-using a timeout mechanism.
\item We propose a dynamic timeout value prediction scheme to provide proper timeout value recommendation based on the historical execution time of the timeout affected function.
\item \blue{We have conducted an empirical study over 91 real production timeout bugs to quantify the coverage of our fixing strategy. TFix$^+$ can fix 79\% bugs completely. We have implemented a prototype of TFix$^+$ and conducted experimental evaluation over 16 real world reproduced timeout bugs. The results show that TFix$^+$ can successfully fix 15 out of 16 tested bugs and provide correct timeout configurations in seconds.}
\end{itemize}

The rest of the paper is organized as follows. Section~\ref{sec:design} describes design details. 
Section~\ref{sec:experiment} presents the experimental evaluation. 
Section~\ref{sec:limitation} discusses the limitation of TFix$^+$. 
Section~\ref{sec:relatedwork} discusses related work.
Finally, the paper concludes in Section~\ref{sec:conclusion}.

\section{System Design}
\label{sec:design}

\begin{figure}[!t]
\centering
\includegraphics[width=\linewidth]{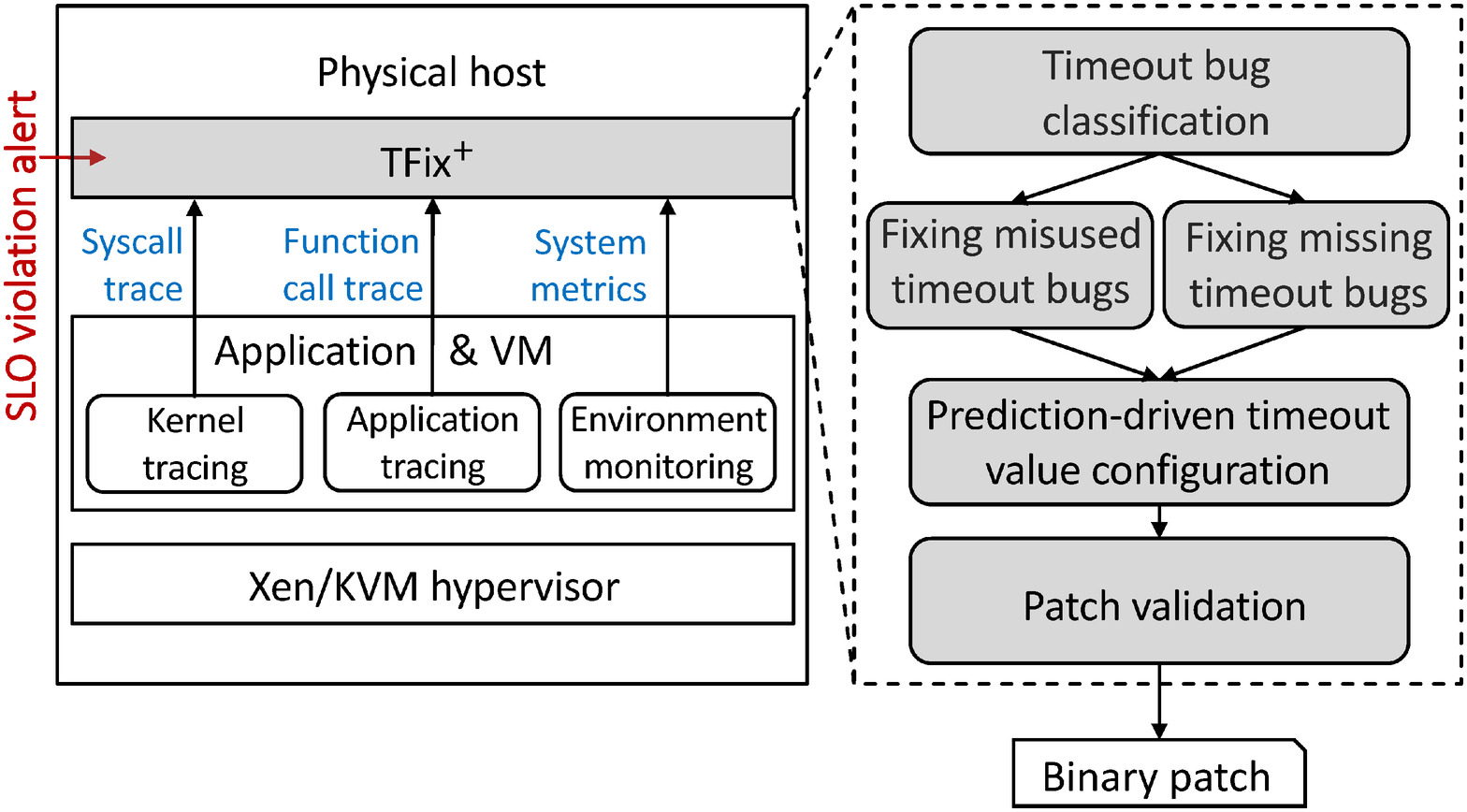}
\caption{The architecture of TFix$^+$.}
\label{fig:tfixArch}
\end{figure}

In this section, we present the design details of the TFix$^+$ system. 
We first provide an overview of TFix$^+$, \blue{followed by the fixing strategies of misused and missing timeout bugs, respectively.}
Next, we describe our prediction-driven timeout value configuration and the patch validation. 

\subsection{Approach Overview}
As shown in Figure~\ref{fig:tfixArch},
when a server system experiences software hang or performance slowdown, TFix$^+$ leverages 
TScope~\cite{hetscope} to identify whether the anomaly is caused by a timeout bug after analyzing
a window of system call trace collected by the kernel tracing module LTTng~\cite{desnoyers2006lttng}.
If TScope confirms that the performance anomaly is caused by a timeout bug, TFix$^+$ is triggered to conduct further drill-down analysis. TFix$^+$ first performs timeout bug classification to determine whether the timeout bug is caused by missing timeout check or incorrectly using timeout \blue{variables}, which is described in our previous work~\cite{timeoutfix}. 

If a misused timeout bug is confirmed, \blue{TFix$^+$ identifies those timeout-affected functions by checking the abnormality of their execution time and frequency from application function traces} (Section \ref{sec:function}). 
TFix$^+$ further \blue{checks whether the timeout-affected functions use a too large or a too small timeout value when the bug happens.} (Section \ref{sec:differentiate}).
\blue{For both cases, TFix$^+$ uses corresponding checking conditions to pinpoint the misused timeout variables associated the timeout-affected functions through static taint analysis} (Section \ref{sec:variable}). 

If a missing timeout bug is confirmed, TFix$^+$ identifies \blue{the root cause as either an infinite loop or a blocking function call} by examining the stack traces (Section \ref{sec:triggeringpoint}). TFix$^+$ then adopts different patching strategies to add timeout mechanisms, \blue{either inside the loop body or surrounding the blocking function call} (Section \ref{sec:addtimeout}).

Lastly, TFix$^+$ produces prediction-driven timeout value configuration \blue{to adapt to the user runtime environment automatically based on historical execution data} and performs dynamic validations over the generated patch (Section \ref{sec:recommendation}). 
The whole drill-down bug diagnosis protocol is executed automatically without requiring any human intervention. 
We will describe each component in detail in the following subsections.  

\subsection{Misused Timeout Bug Fixing}
In this subsection, we discuss how TFix$^+$ identifies the timeout variable for a misused timeout bug. TFix$^+$ first identifies the timeout affected function, then differentiates whether the bug is caused by a too large or a too small timeout value. \blue{For each case, TFix$^+$ takes different matching criteria to identify the misused timeout variable via static taint analysis.}

\subsubsection{Timeout Affected Function Identification}
\label{sec:function}

After classifying a detected bug as a misused timeout bug~\cite{timeoutfix}, 
TFix$^+$ identifies the timeout affected functions by checking the abnormality of the functions' execution time and frequency. To achieve this goal, TFix$^+$ leverages a commonly used application performance tracing tool, i.e., Google's Dapper framework~\cite{sigelman2010dapper}. Dapper allows us to trace the beginning and ending timestamps of all function calls and the control flow graph for the diagnosed bug. 
We choose Dapper tracing tool because it supports distributed systems and incurs low runtime overhead to production systems. The existing implementations of Dapper tracing can only be applied to RPC related functions. 
TFix$^+$ augments the Dapper tracing tool to support all timeout related functions, such as IPC functions and synchronization functions~\cite{timeoutfix}.

\begin{figure}[!t]
\centering
\begin{small}
\begin{lstlisting}[frame=single, style=baseC, basicstyle=\notsotiny\ttfamily]
{"i":"1b1bdfddac521ce8", "s":"df4646ae00070999", "b":1543260568612, 
 "e":1543260568654, "r":"RunJar", "p":["84d19776da97fe78"],
 "d":"org.apache.hadoop.hdfs.protocol.ClientProtocol.getDatanodeReport"}
\end{lstlisting}
\vspace*{-1\baselineskip}
\caption{A trace example of Dapper.}
\label{fig:dappertrace}
\end{small}
\end{figure}

After retrieving a Dapper trace for a target bug, 
we first extract the execution time and frequency of all the functions invoked when the bug happens.
Specifically,
we calculate the frequency of each function by simply counting how many times it is invoked in the Dapper trace. 
We calculate the execution time of each function by subtracting the beginning time from the ending time.
Figure~\ref{fig:dappertrace} shows a Dapper trace example with
various labels indicating different information. 
Among them, ``b'' and ``e'' indicate the beginning timestamp and the ending timestamp of a function, respectively. 
``d'' represents the function name and ``r'' represents the process name. \red{We identify the timeout affected functions by
checking the abnormality in the functions' execution time
and frequency.}

\subsubsection{Misused Timeout Bug Differentiation}
\label{sec:differentiate}

\blue{TFix$^+$ checks whether timeout affection functions use a too large or a too small timeout value based on the following rationales.}
If the timeout value is set to be too large, the execution time of the timeout-affected functions become longer than their normal execution time, \blue{causing system hang or slowdown}.
In contrast, if the timeout value is set to be too small, the system will experience repeated failures due to frequent timeout. 
Therefore, the frequency of the timeout-affected functions become higher than their normal execution frequency.

\blue{The HBase-13647 and HBase-6684 bugs show the examples where the timeout value is set too large, i.e., the RPC connection timeout is misconfigured as {\tt Integer.MAX\_VALUE}.
Under normal state, 
the HBase client can successfully exchange messages with the HBase server (e.g., HMaster, RegionServer) within tens of seconds. 
However, when the HBase server fails, the HBase client hangs for about 24 days,
significantly increasing the execution time of the HBase client's RPC function. 
TFix$^+$ identified the RPC function as the timeout affected function which uses a too large timeout value.}

\blue{The HDFS-4301 bug shows the example where the timeout value is set too small.} 
\red{The timeout value is set to 60 seconds. During normal run, NameNode transfers file system image to the Secondary NameNode within 60 seconds. However, when the image size is too large or network congestion occurs, the transmission process cannot be finished in 60 seconds. Under this circumstance, the system experiences repeated timeout failures because the system retries the image transfer but fails each time. TFix$^+$ identified the {\tt doGetUrl} function as the timeout affected function because their invocation frequencies significantly increase. Besides, the execution time of {\tt doGetUrl} function is nearly the same as the timeout variable value setting (i.e., 60 seconds).} 

\subsubsection{Misused Timeout Variable Identification}
\label{sec:variable}

\begin{figure}[!t]
\centering
\begin{small}
\hfill
\begin{minipage}{.465\textwidth}
\begin{lstlisting}[frame=single, style=baseJava]
//hdfs-site.xml                HDFS-4301(v2.0.3-alpha)
1327 <property>
1328  <name>(*@\tikzmark{aLeft}{}@*)@BG@dfs.image.transfer.timeout@BG@(*@\tikzmark{aRight}{}@*)</name>
1329  <value>60000</value>
      ...
1336 </property>
                             /* (*@\tikzmark{bLeft}{}@*)tainted variables(*@\tikzmark{bRight}{}@*) */
//DFSConfigKeys class
862 public static final String 
863  DFS_IMAGE_TRANSFER_TIMEOUT_KEY 
864   = "dfs.image.transfer.timeout";
865 public static final int 
866  (*@\tikzmark{cLeft}{}@*)@BG@DFS_IMAGE_TRANSFER_TIMEOUT_DEFAULT@BG@(*@\tikzmark{cRight}{}@*) = 60 * 1000;

                    /* (*@\tikzmark{dLeft}{}@*)timeout affected function(*@\tikzmark{dRight}{}@*) */
//TransferFsImage class  
258 public static...(*@\tikzmark{eLeft}{}@*)@BG@doGetUrl@BG@(*@\tikzmark{eRight}{}@*)(...) throws IOException {        
     (*@\tikzmark{fLeft}{}@*)/* timeout variable */(*@\tikzmark{fRight}{}@*)
     ...
271  (*@\tikzmark{gLeft}{}@*)@BG@timeout@BG@(*@\tikzmark{gRight}{}@*) = conf.getInt(
272   DFSConfigKeys.DFS_IMAGE_TRANSFER_TIMEOUT_KEY,
273   DFSConfigKeys.DFS_IMAGE_TRANSFER_TIMEOUT_DEFAULT);
     ...            
277  connection.setReadTimeout(@BG@timeout@BG@);
     ... 
319  InputStream stream = connection.getInputStream();
     ...
358  num = stream.read(buf);
     ... 
401 }
\end{lstlisting}
\end{minipage}
\hfill
\begin{tikzpicture}[overlay,remember picture]
    \renewcommand{\VerticalShiftForArrows}{0.0em,0.0ex}
    \DrawBelowROverArrows[red,solid]{a}{b}
    \DrawOverRBelowArrows[red,solid]{c}{b}
    \DrawOverRBelowArrows[red,solid]{e}{d}
    \DrawOverRBelowArrows[red,solid]{g}{f}
\end{tikzpicture}
\vspace*{-1\baselineskip}
\caption{TFix$^+$ uses the static taint analysis to identify the misused timeout variable for the HDFS-4301 bug.
}
\label{fig:tainttracking}
\end{small}
\end{figure}

\blue{TFix$^+$ correlates the pinpointed functions with different timeout variables included in the system configuration files in order to identify 
the specific misused timeout variables that are attributing to the timeout bug.}


To localize which timeout variable is used when the bug happens, 
we first retrieve all the timeout variables in the target system.
In large scale distributed systems, 
timeout variables along with other configurable parameters 
are often stored in specific configuration files~\cite{xu2015systems}.
For example, in a Hadoop system, all the configurable variables are defined with default values in configuration files, such as {\tt HConstant} and {\tt DFSConfigKeys} classes. 
These variables' value can be overridden and customized by users in {\tt .xml} configuration files.
Thus, all the variables appear in systems' configuration files and contain ``timeout'' keyword in their names are potentially related to misused timeout bugs.
Next, we taint all these timeout variables using static taint analysis tools~\cite{checker} and conduct data flow dependency analysis on them to extract all related variables.
We then check whether the timeout affected functions use the timeout related variables.
If a timeout affected function $f$ uses a timeout variable $v_t$, 
we consider $v_t$ as a misused timeout variable candidate.
We then compare the execution time of $f$ with the value of $v_t$ \blue{to identify the misused timeout values}. 

If the bug is caused by a too large value, the execution time of $f$ should be no larger than the value of $v_t$, considering the system experiences a long delay and we may terminate the system tracing earlier. If the bug is caused by a too small value, we consider $v_t$ as the misused timeout variable if the execution time of $f$ matches the value of $v_t$.

For example, Figure~\ref{fig:tainttracking} shows how TFix$^+$ uses the static taint analysis to identify the misused timeout variable for the HDFS-4301 bug.
In this bug, the default timeout value is set to 60 seconds in {\tt DFS\_IMAGE\_TRANSFER\_TIMEOUT\_DEFAULT} in {\tt DFSConfigKeys.java}. If users configure the timeout variable {\tt dfs.image.transfer.timeout} in {\tt hdfs-site.xml}, the system uses the configured value. Otherwise, the system uses the default value.
We annotate both {\tt dfs.image.transfer.timeout} and {\tt DFS\_IMAGE\_TRANSFER\_TIMEOUT\_DEFAULT} as tainted. 
After applying static taint analysis, we find that the timeout affected function {\tt doGetUrl} uses both tainted variables at line \#271-273. Since the user configures the value of {\tt dfs.image.transfer.timeout} in {\tt hdfs-site.xml}, 
we determine that the misused timeout variable is {\tt dfs.image.transfer.timeout}.
We also perform cross validation between the timeout variable value and the execution time of the timeout affected function to confirm whether our timeout variable identification is accurate.

\subsection{Missing Timeout Bug Fixing}
\label{sec:missing}
\blue{In this subsection, we describe how TFix$^+$ identifies the root cause of missing timeout bugs as either an infinite loop or a blocking function call and inserts proper timeout mechanisms to fix the missing timeout bugs.}

\subsubsection{Root Cause Function Pinpointing}
\label{sec:triggeringpoint}



\begin{figure}[!t]
\centering
\begin{lstlisting}[frame=single, style=baseJava, basicstyle=\notsotiny\ttfamily]
//Stack trace dump 1 at time 12:11:30
"main" #1 prio=5 os_prio=0 tid=... nid=... runnable [...]
java.lang.Thread.State: RUNNABLE
 at java.io.FileOutputStream.writeBytes(Native Method)
 at java.io.FileOutputStream.write(FileOutputStream.java:326)
 ...
 at YarnClientImpl.(*@\tikzmark{aLeft}{}@*)@BG@submitApplication@BG@(*@\tikzmark{aRight}{}@*)(buggycode.java:(*@\tikzmark{bLeft}{}@*)@BG@71@BG@(*@\tikzmark{bRight}{}@*))
 at testcode.main(testcode.java:51)
                              /*(*@\tikzmark{eLeft}{}@*)same function(*@\tikzmark{eRight}{}@*)*/ /*(*@\tikzmark{fLeft}{}@*)different lines(*@\tikzmark{fRight}{}@*)*/
//Stack trace dump 2 at time 12:11:31
"main" #1 prio=5 os_prio=0 tid=... nid=... runnable [...]
java.lang.Thread.State: RUNNABLE
 at java.lang.Thread.sleep(Native Method)
 at YarnClientImpl.(*@\tikzmark{cLeft}{}@*)@BG@submitApplication@BG@(*@\tikzmark{cRight}{}@*)(buggycode.java:(*@\tikzmark{dLeft}{}@*)@BG@75@BG@(*@\tikzmark{dRight}{}@*))
 at testcode.main(testcode.java:51)
\end{lstlisting}
\hfill
\begin{tikzpicture}[overlay,remember picture]
    \renewcommand{\VerticalShiftForArrows}{0.0em,0.0ex}
    \DrawBelowROverArrows[red,solid]{a}{e}
    \DrawOverRBelowArrows[red,solid]{c}{e}
    \DrawBelowROverArrows[red,solid]{b}{f}
    \DrawOverRBelowArrows[red,solid]{d}{f}
\end{tikzpicture}
\vspace*{-1.5\baselineskip}
\caption{A stack trace snippet of an infinite loop bug, i.e., Yarn-1630 (v2.2.0) bug.}
\label{fig:stacktrace}
\end{figure}

\begin{figure}[!t]
\centering
\begin{lstlisting}[frame=single, style=baseJava, basicstyle=\notsotiny\ttfamily]
//Stack trace dump 1 at time 10:11:30
"main" #1 prio=5 os_prio=0 tid=... nid=... runnable [...]
java.lang.Thread.State: RUNNABLE
 at JobEndNotifier.(*@\tikzmark{aLeft}{}@*)@BG@httpNotification@BG@(*@\tikzmark{aRight}{}@*)(JobEndNotifier.java:(*@\tikzmark{bLeft}{}@*)@BG@138@BG@(*@\tikzmark{bRight}{}@*))
 at JobEndNotifier.localRunnerNotification(JobEndNotifier.java:148)
 at TestJobEndNotifier.main(TestJobEndNotifier.java:139)
                               /*(*@\tikzmark{eLeft}{}@*)same function(*@\tikzmark{eRight}{}@*)*/      /*(*@\tikzmark{fLeft}{}@*)same line(*@\tikzmark{fRight}{}@*)*/
//Stack trace dump 2 at time 10:11:31
"main" #1 prio=5 os_prio=0 tid=... nid=... runnable [...]
java.lang.Thread.State: RUNNABLE
 at JobEndNotifier.(*@\tikzmark{cLeft}{}@*)@BG@httpNotification@BG@(*@\tikzmark{cRight}{}@*)(JobEndNotifier.java:(*@\tikzmark{dLeft}{}@*)@BG@138@BG@(*@\tikzmark{dRight}{}@*))
 at JobEndNotifier.localRunnerNotification(JobEndNotifier.java:148)
 at TestJobEndNotifier.main(TestJobEndNotifier.java:139)
\end{lstlisting}
\hfill
\begin{tikzpicture}[overlay,remember picture]
    \renewcommand{\VerticalShiftForArrows}{0.0em,0.0ex}
    \DrawBelowROverArrows[red,solid]{a}{e}
    \DrawOverRBelowArrows[red,solid]{c}{e}
    \DrawBelowROverArrows[red,solid]{b}{f}
    \DrawOverRBelowArrows[red,solid]{d}{f}
\end{tikzpicture}
\vspace*{-1.5\baselineskip}
\caption{A stack trace snippet of a blocking bug, i.e., Mapreduce-5066 bug.}
\label{fig:stacktrace2}
\end{figure}

Our previous study~\cite{timeoutstudy} shows that missing timeout usually causes system stuck at a certain state, \blue{i.e, repeating on executing a block of code inside of an infinite loop or endlessly waiting for an asynchronous function call to finish.} 
\blue{
We use the stack trace analysis to pinpoint the hanging functions with infinite loops and the blocking function calls.}

After a timeout bug is classified as missing timeout, TFix$^+$ dumps \blue{the stack trace of the running system continuously.} 
Figure~\ref{fig:stacktrace} shows an example of the dumped stack trace on a process. 
The trace contains each thread's status, (e.g., runnable or waiting) and the detailed information of each Java function, (e.g., function name and the line number). The per-thread trace also shows the explicit control flow that for the two contingent functions, the top function is the callee of the bottom function.

To localize the hang\blue{/blocking} functions, TFix$^+$ first extracts the common function(s) in multiple traces. 
\blue{This is because hang and blocking are long-term stable status. The same function call appears in all the dumped traces, indicating the manifestation of the hang or blocking.}
TFix$^+$ then extracts the inner-most common functions as the candidate root cause functions.


\blue{TFix$^+$ pinpoints the root cause functions after filtering out those candidate functions triggered by long-running background processes, e.g., the heartbeat process,
by analyzing an interrupted application performance trace using Dapper~\cite{sigelman2010dapper}.} 
\blue{Specifically, TFix$^+$ forces the hang/blocking function to terminate abruptly after a period of tracing time. 
TFix$^+$ then localizes the root cause function based on the following criteria:}
1) the function's execution time is larger than its normal execution time; 
2) the function's start time coincides with the time when the performance alert is reported by the timeout bug detection tool such as TScope~\cite{hetscope}, 
and 3) the function's ending time is the termination time forced by TFix$^+$. 
For example, in the MapReduce-5066 bug, the candidate root cause functions are {\tt Listener.run()}, {\tt Responder.doRunLoop()}, {\tt Reader.run()}, and {\tt JobEndNotifier.httpNotification()}. However, the first three functions are triggered by background processes, \blue{thus are not root cause functions}. 
\blue{
TFix$^+$ excludes those three functions because their execution time is not longer than that in normal runs and their start time is much earlier than the bug detection time.
After pruning, TFix$^+$ identifies the {\tt JobEndNotifier.httpNotification()} function as the root cause function.}

\blue{Next, TFix$^+$ compares the line numbers of the inner-most common functions in different traces to identify the bugs as infinite loops or blocking function calls.
For infinite loops where a block of code executes repeatedly, the line numbers of the candidate root cause function in each trace should be different, as shown in Figure~\ref{fig:stacktrace}.
For blocking function calls, the line numbers of the candidate root cause function are always the same, as shown in Figure~\ref{fig:stacktrace2}.
}

\subsubsection{Adding Timeout Mechanism}
\label{sec:addtimeout}
	\begin{figure}[!t]
	\centering
	\begin{lstlisting}[escapechar=|, frame=single, style=baseJava, tabsize=1]					  
//EditLogTailer.java              HDFS-4176(v2.0.2-alpha)
	255 private void triggerActiveLogRoll() {
						...
	257  try {
	258@RED@-@RED@  getActiveNodeProxy().rollEditLog();
	   (*@\tikzmark{aLeft}{}@*)@RED@+@RED@(*@\tikzmark{aRight}{}@*)  @BG@rollEditLogWithTimeout();@BG@
						 ...
	260  } catch (@BG@IOException@BG@ ioe) {
						 ...
	262 }}

@RED@+@RED@private Configuration conf = new Configuration();
@RED@+@RED@private String @BG@ROLLEDITS_TIMEOUT_KEY@BG@ = "rolledits.timeout";
@RED@+@RED@private int @BG@timeout@BG@ =conf.getInt(ROLLEDITS_TIMEOUT_KEY);			
//a callable thread with timeout setting
(*@\tikzmark{bLeft}{}@*)@RED@+@RED@(*@\tikzmark{bRight}{}@*)public void rollEditLogWithTimeout() throws @BG@IOException@BG@{   
@RED@+@RED@ ExecutorService executor = 
@RED@+@RED@        Executors.newSingleThreadExecutor();
@RED@+@RED@ Callable<Void> callable = new Callable<Void>(){
@RED@+@RED@  @Override
@RED@+@RED@  public Void call() throws IOException {
@RED@+@RED@   return getActiveNodeProxy().rollEditLog();
@RED@+@RED@ }};
@RED@+@RED@ Future<Void> future = executor.submit(callable);
@RED@+@RED@ try {            //timeout setting
@RED@+@RED@		@BG@future.get(timeout, TimeUnit.MILLISECONDS);@BG@
@RED@+@RED@ } catch (Exception e) {
@RED@+@RED@  future.cancel(true);      //acceptable exception
@RED@+@RED@		@BG@throw new IOException;@BG@
@RED@+@RED@ } finally { executor.shutdown(); }
@RED@+@RED@ return null;
@RED@+@RED@}

\end{lstlisting}
\begin{tikzpicture}[overlay,remember picture]
				\renewcommand{\VerticalShiftForArrows}{0.0em,+0.2ex}
				\renewcommand{\HorizonShiftForArrows}{+2.8em,0ex}
				\renewcommand{\HorizonShiftForArrowsCustm}{-2.2em,+0.7ex}
				\renewcommand{\VerticalShiftForArrowsCustm}{0em,+0.2ex}
				\renewcommand{\VerticalShiftForArrowss}{0em,+0.3ex}
				\renewcommand{\VerticalShiftForArrowsss}{+1.5em,+4.8ex}
				\renewcommand{\VerticalShiftForArrowsCs}{-2em,+4ex}
				\renewcommand{\VerticalShiftForArrowsRtoLEnd}{-3.2em,2.4ex}
				\renewcommand{\VerticalShiftForArrowsRtoRStart}{3.2em,2.0ex}
				\DrawArrowsLtoL[azure, in=100, out = -140, solid]{a}{b}
\end{tikzpicture}
\vspace*{-1.5\baselineskip}
\caption{
Code snippet of HDFS-4176 bug. 
When the bug happens, {\tt getActiveNodeProxy().rollEditLog()} on line 258 is stuck.
``{\color{red}-}'' means deleted code and 
``{\color{red}+}'' means added code, representing the patch transformed from TFix$^+$'s instrumented binary code. 
}
\label{fig:HDFS-4176}
\end{figure}

TFix$^+$ adopts different strategies to add the timeout mechanisms for the system hangs on 
\blue{an infinite loop or a blocking function call.}

\blue{For missing timeout induced infinite loops,} 
we add a timeout \blue{checking in the loop body} to ensure the system jump out the loop after the \blue{pre-defined time elapses}.
Specifically, TFix$^+$ adds a configurable timeout variable and marks down the starting time point before the loop is executed. In every loop iteration, TFix$^+$ checks whether the elapsed time exceeds the configured timeout value using an {\tt if} branch. If the checking result is yes, TFix$^+$ breaks out of the loop by throwing an exception.
For example, the Yarn-1630 (v2.2.0) bug, shown in Figure~\ref{fig:Yarn-1630}, and described in Section~\ref{sec:motivating} falls into this category. 

\blue{For missing timeout induced blocking function calls, 
TFix$^+$ reuses existing timeout settings if applicable,
or adds new timeout mechanisms surrounding the blocking function call.}
\red{Existing timeout mechanisms can be provided by overloaded functions or existing methods inside the same class. Overloaded methods are easily to identify.}
For example, if the function is stuck on {\tt Socket.connect(SocketAddress endpoint)}, TFix$^+$ examines all the methods in {\tt Socket} class and finds the \blue{overloaded function} {\tt Socket.connect(SocketAddress endpoint, int timeout)} \blue{with timeout settings}. TFix$^+$ replaces the original one with the supplementary method to add a timeout setting. \red{Besides the overloaded functions, TFix$^+$ examines all the methods of the invoked class and searches the keywords ``set'' and ``timeout'' in all the methods' names. If the method name contains both keywords, we consider it as a candidate timeout setting. TFix$^+$ calls the method to add the timeout setting to the function call. For example, for HDFS-3180 bug, the system hangs on {\tt URLConnection.connect()} function. We search all the methods inside the {\tt URLConnection} and find {\tt connection.setConnectTimeout()} and {\tt connection.setReadTimeout()} methods. TFix$^+$ leverages the two existing methods to add timeout mechanisms.} 

If there is no existing configured timeout mechanisms, TFix$^+$ puts the function in a callable thread and leverages {\tt ExecutorService} and {\tt Future} interfaces provided by standard Java library to monitor and manager the execution of callable threads. For example, Figure~\ref{fig:HDFS-4176} shows the patch of HDFS-4176 bug. When the bug occurs, the system hangs on the {\tt getActiveNodeProxy().rollEditLog()} function. TFix$^+$ does not find any pre-configured timeout mechanisms, so TFix$^+$ replaces the {\tt getActiveNodeProxy().rollEditLog()} function with {\tt rollEditLogWithTimeout()} function and puts the {\tt getActiveNodeProxy().rollEditLog()} function into a callable thread. TFix$^+$ configures the timeout mechanism using {\tt future.get(timeout, TimeUnit.MILLISECONDS)} method to terminate the callable thread when the elapsed time exceeds the value of {\tt timeout}.  

\subsection{Prediction-driven Timeout Value Configuration and Patch Validation}
\label{sec:recommendation}

After pinpointing the misused timeout variable or adding a configurable timeout variable, TFix$^+$ recommends a proper timeout variable value, to fix the timeout bug. During the normal run, a correct timeout value should be consistent with the expected execution time of the timeout affected function. This runtime execution time depends on the application workload (e.g., input file size) and the runtime available resources (e.g., CPU availability and network bandwidth). TFix$^+$ employs an online timeout value prediction scheme to adapt to dynamic runtime environments. 

To achieve model-driven prediction, we first perform continuous collection of normal execution time of the identified function when the timeout bug is not triggered. We also collect the corresponding runtime system metrics and function input parameters when the function is invoked.
We then employ polynomial regression schemes to establish the mapping from the application workload and runtime environment parameters $S_1, S_2,..., S_k,..., S_N$ to function's execution time $T$. 
\red{The polynomial fitting function is calculated by Equation~\ref{eq:prediction}, where $\beta_j$ is the coefficient for the combinatorial term ${S_1}^{p_1}{S_2}^{p_2}...{S_i}^{p_i}...{S_N}^{p_N}$. Each input parameter $S_i$ has the polynomial degree $p_i$. The sum of the degree $p_i$ should be no larger than $P$ which is the highest polynomial degree. This equation contains all the combinatorial terms constructed by all the input parameters when the polynomial degree is no greater than $P$.}
\begin{equation}
\begin{aligned}
T_r = \sum(\beta_j{S_1}^{p_1}{S_2}^{p_2}...{S_i}^{p_i}...{S_N}^{p_N})  \\
where \quad p_i \geq 0 \quad and \quad \sum_{i=1}^{N}(p_i) \leq P
\end{aligned}
\label{eq:prediction}
\end{equation}


Based on our experimental evaluation results, we set the highest polynomial degree $P$ to 3 to both avoid overfitting and achieve fast online prediction. TFix$^+$ changes the $P$ from 1 to 3 and selects the best fitting model using the least-square error.

\green{In practice, any prediction model can hardly achieve perfect prediction. In order to achieve effective timeout bug fixing, TFix$^+$ includes built-in prediction error handling mechanisms. Particularly, under-estimation error (i.e., predicted timeout value is less than expected timeout value) of the prediction model has serious impact to our solution, which causes our timeout bug fix to fail. Thus, we introduce a padding to avoid the under-estimation error. We calculate the estimated timeout values $T_{est}$ produced by the regression model and measure the exact historical execution time $T_{hist}$ during the normal run. We calculate the relative fitting error, which is $(T_{est}-T_{hist})$ over the exact execution time $T_{hist}$, i.e., $\frac{T_{est}-T_{hist}}{T_{hist}}$. Suppose there are $M$ measured data points, the padding ratio is calculated by Equation~\ref{eq:ratio}.
\begin{equation}
T_{ratio} = 2\times \max_{i \in \{ 1, 2,..., M\}}(\frac{T_{est}^i-T_{hist}^i}{T_{hist}^i})
\label{eq:ratio}
\end{equation}
Our experimental results show that our padding scheme can avoid all under-estimations without imposing too much timeout delay. Finally, TFix$^+$ predicts the timeout value $T_{predict}$ using Equation~\ref{eq:padding}.
\begin{equation}
T_{predict} = T_r\times (1 + T_{ratio})
\label{eq:padding}
\end{equation}
}


\begin{figure*}[t!]
    \centering
    \begin{subfigure}[t]{0.45\textwidth}
        \centering
        \includegraphics[width=0.8\linewidth]{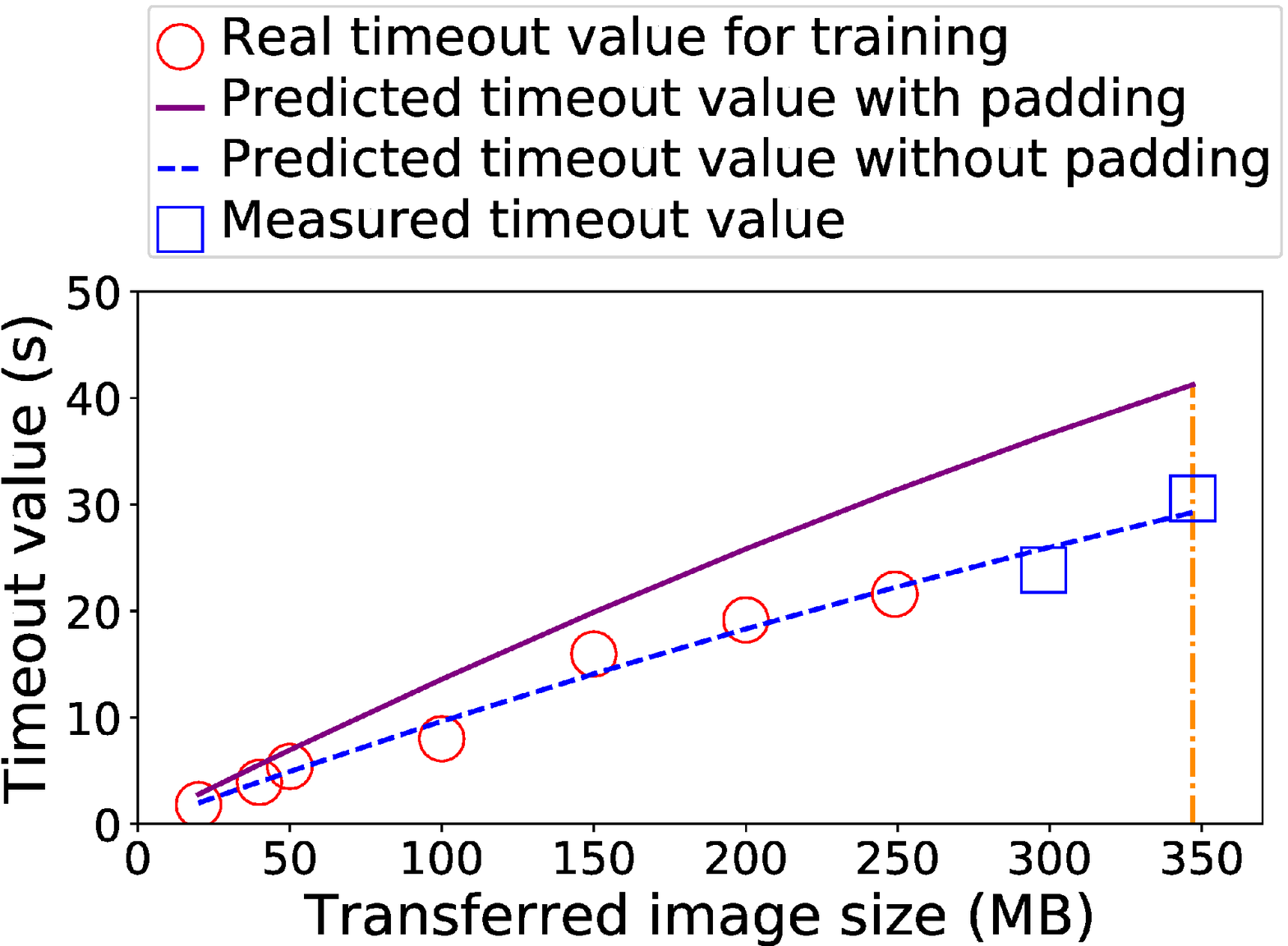}
        \caption{Timeout value prediction based on the function's input (i.e., the transferred image size). }
        \label{fig:prediction_file}
    \end{subfigure}%
    ~
    \hspace{5mm}
    \begin{subfigure}[t]{0.45\textwidth}
        \centering
        \includegraphics[width=0.8\linewidth]{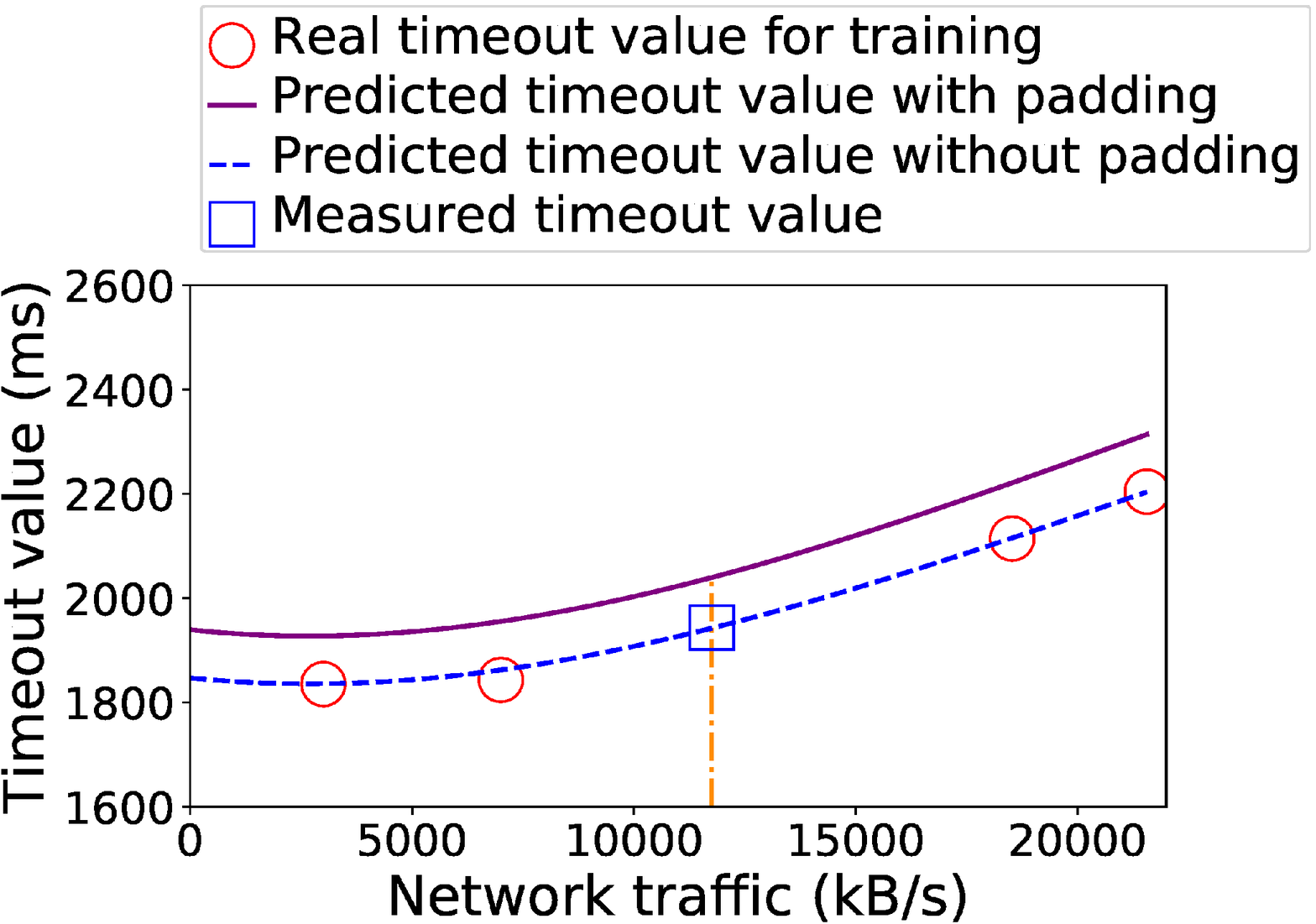}
        \caption{Timeout value prediction based on the network traffic.  }
        \label{fig:prediction_network}
    \end{subfigure}
\caption{Timeout value prediction for the distributed HDFS-4301 bug. The bug occurs when setting improper timeout value on the image transfer between the Primary NameNode and the Secondary NameNode. The Primary NameNode is set up on our lab machine and the Secondary NameNode is set up on an AWS machine. }
\label{fig:prediction_large}
\end{figure*}

The timeout value recommendation considers three different cases: 1) the timeout value is set too large, 2) the timeout value is set too small, and 3) timeout mechanism is missing. 
For the bugs caused by too large timeout values, the regression model is an interpolation model, since the predicted timeout value is within the range of the historical execution time of the timeout affected functions during normal runs. 
For the bugs caused by too small timeout values, the predicted timeout value $T_{predict}$ is beyond the range of the function's execution time $T_{hist}$ in normal runs. In this case, TFix$^+$ builds an extrapolation prediction model to suggest a timeout value. 
For the missing timeout cases, the regression model can be either interpolation or extrapolation model, which depends on the current workload and system environment. \red{For example, 
Figure~\ref{fig:prediction_large} shows the mapping model from the transferred fsimage size and the network traffic to the execution of the function {\tt doGetUrl()} for HDFS-4301 bug. Figure \ref{fig:prediction_file} shows the mapping model of the transferred image size and execution time when we fix the network traffic. Figure \ref{fig:prediction_network} shows the mapping model of the file size and network traffic when we fix the image size to 20 megabytes. Figure \ref{fig:prediction_file} shows an extrapolation model and Figure \ref{fig:prediction_network} shows an interpolation model, respectively.} 

Our online prediction scheme can adapt to different applications, because the predicted value is calculated from the normal run statistics on the specific applications. Our previous work \cite{timeoutfix} adopts a profiling-based heuristic method to provide timeout value recommendation, which might take a long time and high profiling cost to derive the right timeout value configuration.

After the patch is generated, we validate whether the patch is correct or not. A correct patch should fix the bug without introducing new bugs. To check whether the bug is fixed, we apply the patch to the program and re-run the bug triggering test cases. We trigger the bug under the same triggering condition and workload to observe whether the bug occurs again. To check whether our patch introduces new bugs, we run all the test suites provided by the release package (e.g., \cite{testsuite}) to check whether all of them are passed. Normally, a mature system is packaged with test cases to test all the functionalities. For example, Hadoop system is equipped with its JUnit tests to test each class's functions. We determine that TFix$^+$'s patch is valid when it does not influence any other functionalities, that is, all the tests are passed.

\section{Evaluation}
\label{sec:experiment}

In this section, we present our evaluation details. 
We first introduce our evaluation methodology. 
We then present our empirical study results on 91 real world misused or missing timeout bugs.
we then present the experimental evaluation results \blue{on 16 reproduced timeout bugs}. 
Finally we present several case studies.

\blue{All the experiments are conducted in our research lab in a cluster of hosts which are equipped with quad-core Xeon 2.53Ghz CPUs and 16GB memories and running 64-bit Ubuntu v16.04.}
The system call trace is collected using LTTng v2.0.1.
The function call trace is collected using Google's Dapper framework. We adopt existing static taint tracking framework, i.e., Checker~\cite{checker}, to perform misused timeout variable identification. 
We use {\tt jstack} to dump the stack traces for all the Java processes. 
TFix$^+$ is implemented on top of Soot compiler~\cite{soot} to patch binary code.
We use Python's scikit-learn~\cite{scikit} package to build the polynomial regression model to predict timeout values as described in Equation~\ref{eq:prediction} to \ref{eq:padding}.

\subsection{Evaluation Methodology}

\begin{table}[t]
\centering
\caption{System description.}
\scriptsize
\begin{tabular}{|c|l|l|}
\hline
\textbf{System} & \textbf{Setup Mode}  & \multicolumn{1}{c|}{\textbf{Description}}                                                   \\ \hline\hline
Hadoop          & Distributed & \begin{tabular}[c]{@{}l@{}}The utilities and libraries for Hadoop\\ modules\end{tabular}   \\ \hline
HDFS            & Distributed & Hadoop distributed file system                                                              \\ \hline
MapReduce       & Distributed & \begin{tabular}[c]{@{}l@{}}Hadoop big data processing framework    \end{tabular}                                                    \\ \hline
Yarn      & Distributed & \begin{tabular}[c]{@{}l@{}}Hadoop resource management platform    \end{tabular}                                                    \\ \hline
HBase           & Standalone & \begin{tabular}[c]{@{}l@{}}Non-relational, distributed database  \end{tabular}                                                       \\ \hline
Flume           & Standalone & \begin{tabular}[c]{@{}l@{}}Log data collection/aggregation \\ /movement service  \end{tabular}                                                       \\ \hline
\end{tabular}
\label{system_description}
\end{table}

\begin{table*}[]
\centering
\caption{Timeout bug benchmarks.}
\label{benchmark}
\scriptsize
\begin{tabular}{|c|c|c|l|c|c|}
\hline
\textbf{Bug Type}                                                                      & \textbf{Bug ID}                & \textbf{System Version} & \multicolumn{1}{c|}{\textbf{Root Cause}}                                                                                                         & \textbf{Impact}      & \textbf{Workload}   \\ \hline\hline
\multirow{7}{*}{\begin{tabular}[c]{@{}c@{}} too large \\ timeout\end{tabular}} & Hadoop-9106           & v2.0.3-alpha   & ``ipc.client.connect.timeout'' is misconfigured                                                                                         & Slowdown    & Word count \\ \cline{2-6} 
                                                                              & \begin{tabular}[c]{@{}c@{}} Hadoop-11252 \\(v2.6.4) \end{tabular} & v2.6.4         & Timeout is misconfigured for the RPC connection                                                                                         & Hang        & Word count \\ \cline{2-6} 
                                                                              & HDFS-10223            & v2.8.0         & \begin{tabular}[c]{@{}l@{}}Timeout value on setting up the SASL connection \\ is too large\end{tabular}                                 & Hang        & Word count \\ \cline{2-6} 
                                                                              & MapReduce-4089        & v2.7.0         & ``mapreduce.task.timeout'' is set too large                                                                                             & Slowdown    & Word count \\ \cline{2-6} 
                                                                              & Yarn-1630 (v2.3.0)    & v2.3.0         & \begin{tabular}[c]{@{}l@{}}The default value of ``client.application-client-protocol\\ .poll-timeout-ms'' is set too large\end{tabular} & Hang        & Word count \\ \cline{2-6} 
                                                                              & HBase-15645           & v1.3.0         & ``hbase.rpc.timeout'' is ignored                                                                                                        & Slowdown    & YCSB       \\ \cline{2-6} 
                                                                              & HBase-17341           & v1.3.0         & \begin{tabular}[c]{@{}l@{}}Timeout is misconfigured for terminating \\ replication endpoint\end{tabular}                                & Slowdown    & YCSB       \\ \hline
\multirow{4}{*}{\begin{tabular}[c]{@{}c@{}}too small \\ timeout\end{tabular}} & Hadoop-10695          & v2.6.0         & \begin{tabular}[c]{@{}l@{}} Timeout is misconfigured for the KMSClientProvider \end{tabular}                                                                                     & Job failure & Word count \\ \cline{2-6} 
                                                                              & HDFS-4301             & v2.0.3-alpha   & Timeout value on image transfer operation is small                                                                                      & Job failure & Word count \\ \cline{2-6} 
                                                                              & HDFS-9887             & v2.8.0         & \begin{tabular}[c]{@{}l@{}}Timeout value on WebHdfs socket connection is \\ too small\end{tabular}                                      & Job failure & Word count \\ \cline{2-6} 
                                                                              & MapReduce-6263        & v2.7.0         & ``hard-kill-timeout-ms'' is misconfigured                                                                                               & Job failure & Word count \\ \hline
\multirow{5}{*}{\begin{tabular}[c]{@{}c@{}}missing \\ timeout\end{tabular}}   & HDFS-3180             & v2.0.4-alpha   & Socket timeout is missing on WebHDFS connection                                                                                         & Hang        & Word count \\ \cline{2-6} 
                                                                              & HDFS-4176             & v2.0.2-alpha   & \begin{tabular}[c]{@{}l@{}}Timeout is missing when EditLogTailer calls rollEdits \end{tabular}                                                                                  & Hang        & Word count \\ \cline{2-6} 
                                                                              & MapReduce-5066        & v2.0.3-alpha   & Timeout is missing when JobTracker calls a URL                                                                                          & Hang        & Word count \\ \cline{2-6} 
                                                                              & Yarn-1630 (v2.2.0)    & v2.2.0         & \begin{tabular}[c]{@{}l@{}}Timeout is missing for asynchronous polling operations  \end{tabular}                                                                                & Hang        & Word count \\ \cline{2-6} 
                                                                              & Flume-1819            &      v1.3.0          &  Timeout is missing for reading data                                                                                                                                       & Hang        &      Writing log events     \\ \hline
\end{tabular}
\end{table*}

We first conduct an empirical study on 91 real world misused or missing timeout bugs which are collected by our previous study~\cite{timeoutstudy} \blue{to manually evaluate the fixing strategies of TFix$^+$}.
We check each bug to determine whether TFix$^+$ can fix the bug without introducing new problems. 
\blue{We should note that, our benchmarks do not contain concurrency bugs, e.g., race conditions and deadlocks. Even though TFix$^+$ can fix some of them by adding timeout mechanisms, in most cases, fixing concurrency bugs requires specific techniques, e.g., order checking~\cite{jin2012automated}, atomicity checking~\cite{jin2011automated} and etc.}

\blue{To the best of our effort, we have reproduced 16 timeout bugs from six commonly-used open-source systems, listed in Table~\ref{system_description}.}
We set up four systems in the distributed mode to investigate timeout issues occurring on the communications among different nodes.
\blue{The 16 reproduced bugs are listed in Table~\ref{benchmark}, including 11 misused and 5 missing timeout bugs.
They cover timeout problems in both network communications and intra-node coordinations, representing common real-world missing and misused timeout bugs.}


{\bf Workload:} \blue{To replay real-word scenarios, each system runs a specific workload, listed in Table~\ref{benchmark}.}
Specifically, 
for the Hadoop, HDFS, MapReduce and Yarn systems, we run word count job on a 765MB text file. 
For the HBase system, we use the YCSB workload generator to make insertion, query and update operations on a table.
For the Flume system, we write log events to the log collection tool and distribute the logs repeatedly. 
These workloads invoke the timeout affected functions all of our tested timeout bugs. 

{\bf Expected timeout value measurement:} In order to validate the prediction accuracy, we measure the exact expected timeout value during buggy run. We run the same workload under same runtime computing environment without triggering the bug and retrieve the timeout affected function's execution time in the function call trace, which is the exact expected timeout value. 

\subsection{Empirical Study Results}
\label{sec:empirical}

\begin{figure}[!t]
\centering
\includegraphics[width=0.9\linewidth]{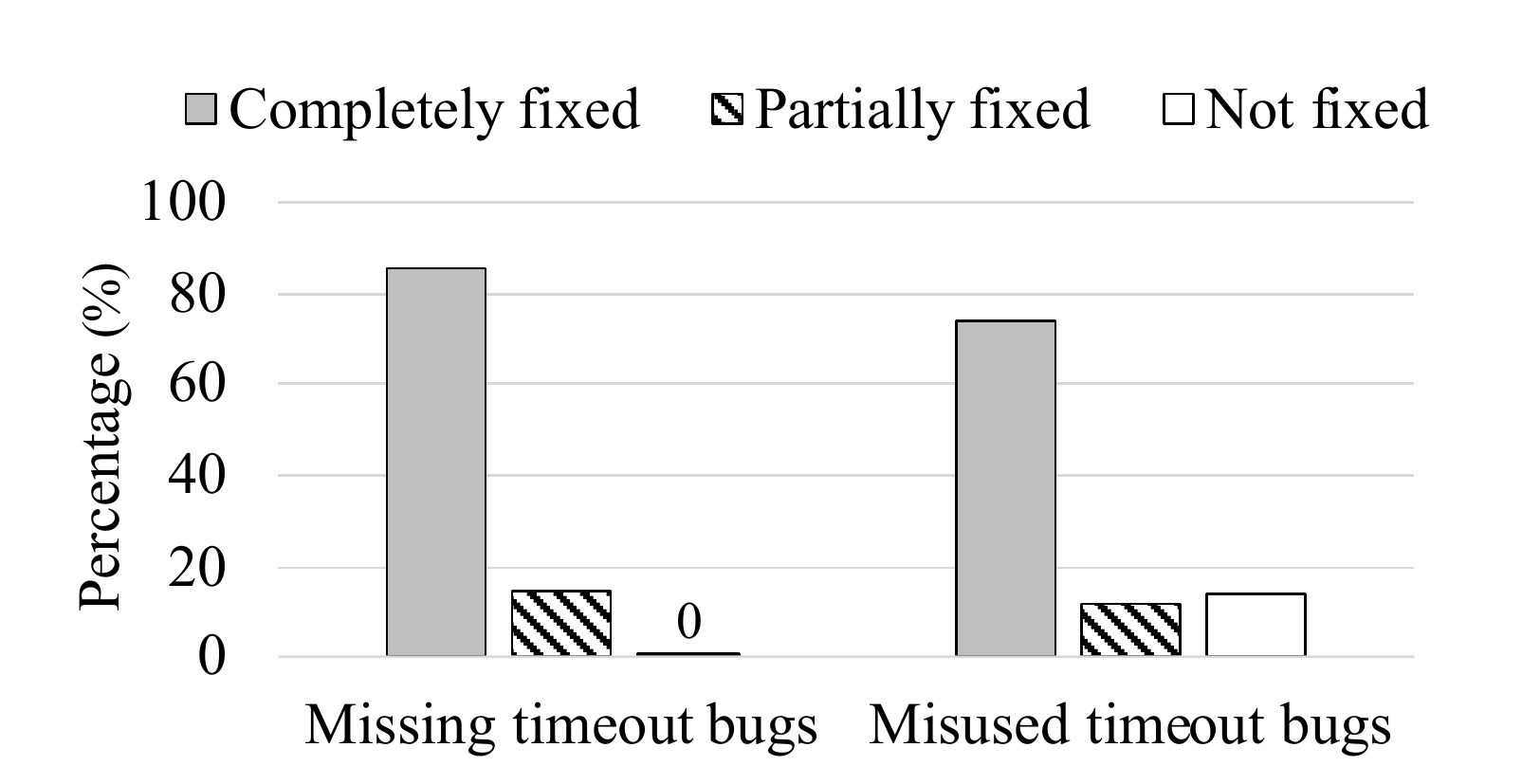}
\caption{Fixing coverage of TFix$^+$.}
\label{fig:study}
\end{figure}

Figure \ref{fig:study} shows the fixing coverage of TFix$^+$ on 91 studied timeout bugs. 
\blue{TFix$^+$ fixes 72 bugs (i.e., 79\%) completely. TFix$^+$ fixes 12 (i.e., 13\%) bugs partially because their fixes require not only proper timeout mechanisms but also retry schemes and program semantic repairs. For the rest seven (i.e., 8\%) bugs which cannot be fixed by TFix$^+$, they are all caused by improper hard-coded timeout values.} 


The seven misused timeout bugs cannot be fixed by TFix$^+$ are caused by hard-coded timeout values, \blue{i.e., no configurable timeout variables.}
Those timeout bugs often occur in early versions of a system, such as Hadoop 0.x version and HBase 0.x version. 
TFix$^+$ cannot identify the \blue{misconfigured} timeout \blue{variables} in our taint analysis since the timeout affected functions do not use any configurable parameter or configuration constant. 
Although TFix$^+$ cannot localize misused timeout variables, TFix$^+$ identifies the bug as a misused timeout bug and pinpoint the timeout affected function. The hard-coded timeout value should be located in the timeout affected function or its callee functions. Developers can perform a cross validation to check whether any hard-coded value equals to the timeout affected function's execution time. Then developers can expose a configurable timeout variable for users.

\subsection{\red{Experimental Results}}
\label{sec:result}
In this subsection, we present the experimental results on 16 reproduced timeout bugs.
we first present the results of fixing misused timeout bugs and fixing missing timeout bugs,
and then we describe the results of timeout value prediction and patch validation. Finally, we present the negative case study to show why TFix$^+$ cannot fix one timeout bug completely.

\subsubsection{Missing Timeout Bugs Fixing Results}




\begin{table}[t]
\centering
\caption{Results of adding timeout mechanisms. 
}
\scriptsize
\begin{tabular}{|c|c|c|}
\hline
\textbf{Bug ID}    & \textbf{Hang Type} & \textbf{\begin{tabular}[c]{@{}l@{}}Missing Timeout Bug \\Fixing Strategy\end{tabular}}                                                                                \\ \hline\hline
HDFS-3180          & Blocking func      & \begin{tabular}[c]{@{}l@{}}Use existing timeout setting\end{tabular}                                       \\ \hline
HDFS-4176          & Blocking func      & \begin{tabular}[c]{@{}l@{}}Leverage {\tt future.get()} method \\of a callable thread\end{tabular}                                                    \\ \hline
MapReduce-5066     & Blocking func      & \begin{tabular}[c]{@{}l@{}}Leverage {\tt future.get()} method \\of a callable thread\end{tabular} \\ \hline
Yarn-1630 (v2.2.0) & Infinite loop               & \begin{tabular}[c]{@{}l@{}}Check whether the elapsed time \\exceeds the pre-defined timeout \\value \end{tabular}                                            \\ \hline
Flume-1819 & Infinite loop              & \begin{tabular}[c]{@{}l@{}}Check whether the elapsed time \\exceeds the pre-defined timeout \\value \end{tabular}                                            \\ \hline
\end{tabular}
\label{addtimeout}
\end{table}

Table~\ref{addtimeout} shows the results of adding timeout mechanisms \blue{on the 5 missing timeout bugs.
TFix$^+$ fixes the HDFS-3180 bug by first identifying that it hangs on a blocking function call and then adding timeout mechanisms using existing timeout-setting functions, i.e., {\tt connection.setConnectTimeout()} and {\tt connection.setReadTimeout()}.
For the HDFS-4176 and MapReduce-5066 bugs, there are no existing timeout mechanisms that can be set on corresponding blocking function calls.
TFix$^+$ fixes them by isolating the blocking functions in callable threads with timeout settings on the {\tt future.get()} method.

TFix$^+$ fixes the Yarn-1630 (v2.2.0) and Flume-1819 bugs by first identifying that they hang on infinite loops and then adding timeout checking on the loop body. Specifically, TFix$^+$ checks whether the elapsed time exceeds the pre-defined timeout value in each loop iteration.
} 

\subsubsection{Prediction-driven Timeout Value Configuration and Patch Validation Results}

\begin{table*}[!t]
\centering
\caption{The fixing result of TFix$^+$. Note that in some bugs, the developers expose the timeout variable for users to configure, without changing the default value. The default value causes bugs under current workload. We mark them as ``manual threshold'' in the table.}
\label{fix_result}
\scriptsize
\begin{tabular}{|c|c|c|c|c|c|c|c|c|}
\hline
 \textbf{Bug Type}                                                                     & \textbf{Bug ID}                & \textbf{Impact}      &  \textbf{\begin{tabular}[c]{@{}c@{}}Buggy\\ Timeout\\ Value\end{tabular}} & \textbf{\begin{tabular}[c]{@{}c@{}}Timeout Value\\ in the Manual \\Patch\end{tabular}} & \textbf{ \begin{tabular}[c]{@{}c@{}}Predicted\\ Timeout Value\end{tabular}} & \textbf{\begin{tabular}[c]{@{}c@{}}Expected\\ Timeout\\ Value\end{tabular}} &  \textbf{Bug Fixed} & \textbf{Diagnosis Time} \\ \hline\hline
\multirow{7}{*}{\begin{tabular}[c]{@{}c@{}}too large\\ timeout\end{tabular}} & Hadoop-9106           & Slowdown    & 20s                                                   & \begin{tabular}[c]{@{}c@{}}manual threshold\end{tabular}                     & 2.80s                              & 1.63s                                                                                      & \cmark        & 3.13s                                                                                  \\ \cline{2-9} 
                                                                             & Hadoop-11252 (v2.6.4) & Hang        & Infinity                                              & 60s                               & 0.10s                                      & 0.05s                                                                                                                           & \cmark      & 3.04s                                                                                    \\ \cline{2-9} 
                                                                             & HDFS-10223            & Hang        & 2h                                                    & 60s                                    & 0.11s                                  & 0.09s                                                                                                                         & \cmark        & 3.16s                                                                                  \\ \cline{2-9} 
                                                                             & MapReduce-4089        & Slowdown    & 5min                                                  & \begin{tabular}[c]{@{}c@{}}manual threshold\end{tabular}        & 3.86s    & 3.50s                                                                                                                             & \cmark     & 2.26s                                                                                     \\ \cline{2-9} 
                                                                             & Yarn-1630 (v2.3.0)            & Hang        & Infinity                                              & \begin{tabular}[c]{@{}c@{}}manual threshold\end{tabular}                  & 1.11s   & 1.10s                                                                                                                    & \cmark     & 1.78s                                                                                     \\ \cline{2-9} 
                                                                             & HBase-15645           & Slowdown        & 20min                                                 & 1min                         & 3.47s                                         & 1.98s                                                                                                                            & \cmark           & 2.60s                                                                               \\ \cline{2-9} 
                                                                             & HBase-17341           & Slowdown    & 5min                                                  & \begin{tabular}[c]{@{}c@{}}manual threshold\end{tabular}                     & 0.13s      & 0.11s                                                                                                              & \cmark      & 1.87s                                                                                    \\ \hline
\multirow{4}{*}{\begin{tabular}[c]{@{}c@{}}too small\\ timeout\end{tabular}} & Hadoop-10695          & Job failure & 60s                                                   & \begin{tabular}[c]{@{}c@{}}manual threshold\end{tabular}        & 137.99s       & 135.05s                                                                                                                      & \cmark     & 3.14s                                                                                        \\ \cline{2-9} 
                                                                             & HDFS-4301             & Job failure & 60s                                                   & \begin{tabular}[c]{@{}c@{}}manual threshold\end{tabular}          & 138.68s     & 119.64s                                                                                                                      & \cmark       & 3.30s                                                                                   \\ \cline{2-9} 
                                                                             & HDFS-9887             & Job failure & 60s                                                   & \begin{tabular}[c]{@{}c@{}}manual threshold\end{tabular}        & 87.90s        & 85.39s                                                                                                                       & \cmark        & 3.40s                                                                                  \\ \cline{2-9} 
                                                                             & MapReduce-6263        & Job failure & 10s                                                   & \begin{tabular}[c]{@{}c@{}}manual threshold\end{tabular}                & 13.10s     & 13.05s                                                                                                                  & \cmark      & 2.85s                                                                                     \\ \hline

\multirow{4}{*}{\begin{tabular}[c]{@{}c@{}}missing\\ timeout\end{tabular}} & HDFS-3180         & Hang & --                                                   & \begin{tabular}[c]{@{}c@{}} 1min\end{tabular}          &                 6.40s    &  5.95s                                                                                                       & \cmark     &  2.36s                                                                                         \\ \cline{2-9} 
                                                                             & HDFS-4176             & Hang & --                                                   & \begin{tabular}[c]{@{}c@{}} manual threshold\end{tabular}          & 0.25s    &   0.21s                                                                 & \cmark       &       2.89s                                                                             \\ \cline{2-9} 
                                                                             & MapReduce-5066             & Hang & --                                                    & \begin{tabular}[c]{@{}c@{}}  manual threshold\end{tabular}                                                   & 0.18s                               &  0.16s                                                       & \cmark        &           
                                                                             3.03s                                                                        \\ \cline{2-9} 
                                                                             & Yarn-1630 (v2.2.0)        & Hang & --                                                   & \begin{tabular}[c]{@{}c@{}}  manual threshold\end{tabular}                                                    & 1.13s                          & 1.09s                                                           & \cmark      &   1.48s                                                                                   \\ \cline{2-9} 
                                                                             & Flume-1819        & Hang & --                                                   & \begin{tabular}[c]{@{}c@{}}  3s\end{tabular}                                                    & 30.034s                           &  30.028s                                                            & \xmark      &     5.63s 
                                                                             \\ \hline

\end{tabular}
\end{table*}

Table~\ref{fix_result} shows the bug fixing results by TFix$^+$. 
We list the expected timeout value under current workload for each bug. After adopting TFix$^+$'s patch with the predicted timeout value in the system, we find that 15 out of 16 bugs do not occur anymore under the same workload with all the test suites passed successfully.

We list the timeout values in the bugs' patch files in Table~\ref{fix_result}. 
We observe that the timeout values in the manual patches are not always correct. 11 out of 16 patches of the bugs need user inputs to set manual thresholds. It means that when patching misused timeout bugs, developers usually make the timeout variable configurable for users and set a default value which requires the user to set the timeout value based on his or her own application workloads.
However, it is challenging to make the correct configurations for different production workloads and computing environments, even for experienced engineers.
For example, in the patch of Yarn-1630 (v2.3.0) bug, 
the default value of the {\tt yarn.client.application-client-protocol\\.poll-timeout-ms} variable is configured to be -1 milliseconds, which equals infinity. Developers expose the variable for users to configure. If users do not properly configure the timeout variable, the timeout bug still happens in the patched version.

We also observed that TFix$^+$'s fixing schemes sometimes are different from the manual patches. We use HDFS-4301 bug as an example. In the patch of HDFS-4301 bug, the default value of {\tt dfs.image.transfer.timeout} is still set to 60 seconds, which is identical with the timeout value before patching. However, the patch partitions the fsimage file into multiple chunks of a equal chunk size. The timeout variable limits the maximum transferring time for each chunk. In contrast, TFix$^+$ changes the timeout value to 138.68 seconds, that successfully fix the problem. 

We should note that, the predicted timeout value by TFix$^+$ might be different under different workloads.  
This is our design choice, because a fixed timeout setting cannot handle unexpected workload changes or environment fluctuations.  
For example, in HBase-15645 bug, the misused timeout variable {\tt hbase.client.operation.timeout} defines the time to block a certain table to prevent concurrency issues. 
Since the table size is small for YCSB workload in our evaluation, the predicted value by TFix$^+$ is only 3.47 seconds. 
If we use a large timeout value under the same YCSB workload, the user will still experience a noticeable delay in the system. On the other hand, if we increase the table size, the predicted value will increase correspondingly.

Although TFix$^+$ introduces timeout value over-estimations to avoid too small timeout bugs, the differences between the predicted values and the expected values are no more than 20 seconds for all the tested bugs.

We also list the diagnosis time for each bug in Table~\ref{fix_result}, which is the total execution time for all the fixing components of TFix$^+$. The diagnosis time is within ten seconds, which makes it practical to apply TFix$^+$ in real-world systems. The runtime overhead comes from three tracing modules, i.e., system call tracing, function call tracing and system metric tracing. As discussed in our previous work~\cite{timeoutfix}, function call tracing incurs low overhead because TFix$^+$ only traces a small number of functions (functions related to timeout configuration, network connection and synchronization). Our prototype implementation experiments show that TFix$^+$ incurs less than 3\% runtime overhead to the monitored system.

\subsubsection{Negative Case Study}
\label{sec:negative}
As shown in Table~\ref{fix_result}, Flume-1819 bug cannot be fixed by TFix$^+$ since not all the test suites are passed after adopting TFix$^+$'s patch. Flume-1819 bug is triggered when the system hangs on an infinite loop to read the data into channels. To fix the bug, a timeout mechanism is needed to flush the cache data into disk when the bug is triggered. Otherwise, the system experiences data loss.  
TFix$^+$ can partially fix the bug by enabling the system to jump out of the infinite loop. However, in order to fix the bug completely, we need to trigger cache data flushing, which requires application specific knowledge.
In addition, TFix$^+$ provides a proper value for the timeout mechanism, which complements the manual patch. As shown in Table~\ref{fix_result}, the default value of the manual patch is not applicable under current workload. The expected timeout value (i.e., 30 seconds) is significantly larger than the one (i.e., 3 seconds) in the manual patch. 

\subsection{Case Study}
In our previous study paper \cite{timeoutstudy}, we further divide the misused and missing timeout bugs into multiple subcategories based on the root causes, e.g., incorrectly reused timeout values, ignored timeout values, missing timeout for network communication and missing timeout for synchronization. We list several cases to show how TFix$^+$ fixes bugs belonging to different subcategories. 

{\bf HBase-16556 (misused):} this bug is caused by reusing {\tt rpcTimeout} in {\tt get()}, {\tt delete()}
and {\tt existsAll()} functions. When this bug occurs, {\tt get()} function invokes {\tt batch()} function and invokes {\tt submitAll()} function. {\tt submitAll()} leverages the {\tt rpcTimeout} variable, whose value causes severe performance degradation to the system. TFix$^+$ can identify {\tt submitAll()} function as the timeout affected function and further determine the misused variable {\tt rpcTimeout}. TFix$^+$ then adopts the prediction scheme to suggest a proper value for {\tt rpcTimeout}. The manual patch leverages the timeout mechanism of {\tt batch()} function to introduce a new variable with a recommended value, which has the same effect of TFix$^+$'s bug fix.

{\bf HBase-13104 (misused):} this bug is caused by ignoring a pre-configured timeout variable to adjust the session timeout for the standalone HBase system. Instead, HBase system only uses the configuration constant {\tt ZK\_SESSION\_TIMEOUT} as the session timeout. The constant is defined in the configuration file {\tt HConstants.java} and TFix$^+$ can localize it by performing taint analysis. TFix$^+$ then predicts a proper session timeout value for the constant {\tt ZK\_SESSION\_TIMEOUT} instead of introducing another variable. 

{\bf Hadoop-4659 (missing):} this bug is caused by missing IPC timeout for network communication. TFix$^+$ can identify an infinite loop inside the {\tt waitForProxy()} function. TFix$^+$ then adds a timeout mechanism to jump out of the loop, which is the same as the manual patch.

{\bf Phoenix-2496 (missing):} this bug is caused by missing timeout for synchronization. The bug happens when a query thread on QueryServer holds a read lock to connect with HBase and another shutdown hook thread is acquiring the lock. HBase system cannot respond to QueryServer, due to heavy workload or unexpected events. Then the query thread does not release the lock and the shutdown hook thread hangs. TFix$^+$ identifies the systems hangs on {\tt closeInstance()} inside the {\tt addShutdownHook} thread. Since no pre-configured timeout mechanism is found, TFix$^+$ then leverages {\tt future.get()} to add a timeout mechanism on {\tt closeInstance()} function. The manual patch is the same with TFix$^+$'s patch.

\subsubsection{Negative Case Study}
We describe two cases to show why TFix$^+$ fix the bugs partially or TFix$^+$ cannot fix the bugs.

{\bf HBase-3456 (misused):} this bug is caused by ignoring the configurable timeout variable {\tt ipc.socket.timeout}. HBase client uses a hard-coded 20 seconds as the socket timeout in {\tt NetUtils.connect()} function. TFix$^+$ can identify {\tt setupIOstreams()} inside the {\tt HBaseClient} class as the timeout affected function. However, since no pre-configured timeout variable is used, TFix$^+$ cannot identify the root cause since the 20-second socket timeout is hard-coded. 

\green{{\bf MapReduce-4813 (missing):} this bug is caused by missing timeout for synchronization. The bug occurs when the Application Manager (AM) calls {\tt commitJob} method synchronously during JobImpl state transitions. The JobImpl write lock is held by the {\tt commitJob} thread. If committing the job takes too long, the state transition hangs on acquiring the lock. TFix$^+$ can partially fix the bug by introducing a timeout mechanism to terminate the hanging job commit thread. However, TFix$^+$ cannot reset the JobImpl state. In comparison, the manual patch adds the {\tt COMMIT\_COMPLETED} and {\tt COMMIT\_FAILED} states and enables the job to transit to a proper state after the commit succeeds or fails.}

\section{Limitation}
\label{sec:limitation}

\green{TFix$^+$ fixes both misused timeout bugs and missing timeout bugs which cover 78\% timeout bugs in total. When fixing misused timeout bugs, TFix$^+$ can localize the misused timeout variable if the cloud system uses timeout variables in timeout handling operations. However, we observe that some timeout bugs are caused by hard-coded timeout values. The hard-coded cases are discussed in Section \ref{sec:empirical}. Although TFix$^+$ cannot localize misused timeout variables, TFix$^+$ can correctly identify them as the misused timeout bugs and pinpoint the timeout affected functions. TFix$^+$ provides useful hints to localize the hard-coded values.

TFix$^+$ currently cannot cover the other three categories of timeout bugs, i.e., improper handling, unnecessary timeout, and clock drifting. For the bugs caused by improper handling, we need to add or revise application-specific functions to fix them. 
For the bugs caused by unnecessary timeout, TFix$^+$ can identify the timeout affected functions when the bugs cause performance issues. 
However, whether removing the unnecessary timeout alters the control flow and brings any unwanted effects is out of the scope of this paper. 
For the bugs caused by clock drifting, the fixing strategy is to enhance the synchronization of distributed systems, rather than to modify any timeout mechanisms.}

\red{TFix$^+$ currently only supports cloud systems written in Java since the implemation of fixing misused and missing timeout bugs only works on Java platforms now. 
However, our approach is agnostic to programming languages. TFix$^+$ can be easily extended to support other programming languages by replacing Java specific Dapper, static taint analysis, timeout mechanism instrumentation components with other programming language counterparts. }

\section{Related Work}
\label{sec:relatedwork}

\textbf{Automatic bug fixing.}
Work has also been done to automatically fix different bugs. 
For example, 
AFix~\cite{jin2011automated} and CFix~\cite{jin2012automated} proposed automatic patching strategies for concurrency bugs. 
ClearView~\cite{perkins2009automatically} identified violated invariants from erroneous executions and generated candidate repair patches to change the invariants.
Tian et al.~\cite{tian2012identifying} presented an automatic bug fixing patch identification tool to maintain older stable versions.  
Tufano et al.~\cite{tufano2018empirical} applied an Encoder-Decoder model based on neural networks to mine the existing patches and automatically generate new ones. 
\red{Narya \cite{levy2020predictive} predicted a host failure by adopting domain knowledge and machine learning models and took appropriate actions to mitigate failures.} 
In comparison, our work fixes timeout bugs with the drill-down bug analysis approach that \blue{can not only identify bug root causes but also apply proper timeout fixing strategies.} 

\blue{Our previous work HangFix \cite{he2020hangfix} focused on fixing hang bugs caused by infinite loops and blocking function calls. 
HangFix proposed to resolve some hang problems by inserting a timeout mechanism with a pre-set timeout variable. In contrast, TFix$^+$ focuses on fixing both misued and missing timeout bugs which could cause performance slowdown, job failures, or software hang.  
In HangFix, the newly introduced timeout variable is heuristically configured, thus cannot handle dynamic workloads.
TFix$^+$ provides the prediction-driven timeout variable configuration scheme which can adapt to user runtime environment and dynamic workloads.}
\red{TFix \cite{timeoutfix} is the early version of TFix$^+$, which focuses on only fixing misused timeout bugs covering 47\% timeout bugs. In contrast, TFix$^+$ fixes both misused and missing timeout bugs that cover 78\% timeout bugs. We have conducted extensive empirical study on the fixing coverage of TFix$^+$. Moreover, TFix$^+$ provides a new prediction-driven approach to achieving self-configuring timeout bug fixing. }

\textbf{Tracing-based bug detection and diagnosis.}
\blue{Previous work has extensively used tracing techniques to detect and diagnose bugs.
For example, X-ray~\cite{xray} dynamically instrumented binaries to trace the inputs and outputs of different components and inferred the traces to diagnose performance bugs.}
Chopstix~\cite{bhatia2008lightweight} collected low-level OS events such as CPU utilization and I/O operations online and reconstructed these events offline for troubleshooting \blue{intermittent and non-reproducible} bugs.
REPT~\cite{cui2018rept} utilized hardware traces to reconstruct the program's execution and employed record-and-replay techniques for debugging.
Magpie~\cite{magpie} instrumented middleware and packet transfer
points to record fine-grained system events and correlated
these events to capture the control-flow and resource consumption of each request for debugging.  
\red{CrashTuner \cite{lu2019crashtuner} detected crash-recovery bugs by injecting faults in the test suites and exposing the bugs via log-based program analysis.} 
\red{Gong et al. \cite{gong2020experiences} conducted a study on applying machine learning techniques to detecting malwares from a major Android App market.} 
\blue{Different from TFix$^+$, these tracing techniques do not apply for timeout bug diagnosis or fixing.}

\blue{Tracing techniques have also been used by previous work to detect timeout bugs specifically.}
For example,
TScope~\cite{hetscope} detected timeout bugs using timeout related feature selection and machine-learning based anomaly detection on system call traces. 
SafeTimer~\cite{ma2018accurate} checked whether a timeout bug is caused by network delay or packet delay at the OS level. 
\blue{Those tracing-based timeout detection techniques are complementary to and can be used with TFix$^+$ as an automated system for timeout detection, diagnosis and fixing. In our experiment, we use TScope as the front-end to identify whether an anomaly is caused by timeout.}

\textbf{Configuration bug detection and diagnosis.}
\blue{Different techniques have been used to detect and diagnose misconfiguration issues.}
Violet~\cite{hu2020automated} adopted symbolic execution to detect specious configurations.  
Ctests~\cite{sun2020testing} generated new test cases based on original test suites to detect errors caused by configuration changes. 
SPEX~\cite{xu2013not} studied configuration constraints and exposed potential configuration errors by injecting errors that violate the constraints.
ConfValley~\cite{huang2015confvalley} introduced a new language to define system validation rules and checked configurations against those rules before the application was deployed in production. 
PCheck~\cite{pcheck} analyzed the application source code and automatically emulated the late execution that used configuration values to detect latent configuration errors. 
CODE~\cite{yuan2011context} detected configuration bugs by identifying the abnormal program executions using invariant configuration access rules.
ConfAid~\cite{attariyan2010automating} adopted dynamic taint tracking methods to instrument the binary application and analyzed the information flow to pinpoint the root causes of configuration errors. 
ConfDiagnoser~\cite{zhang2013automated} extracted the control flow of configuration options, instrumented the application code for profiling and analyzed the configuration deviation to detect the erroneous configuration options. 
EnCore~\cite{zhang2014encore} applied machine learning techniques to model the correlation between the configuration settings and the executing environment and correlations between configuration entries, in order to learn and detect configuration bugs. 
In summary, existing approaches focused on detecting misconfiguration issues that caused functional bugs. They cannot be readily applied to detecting performance issues caused by misconfigured variables that are triggered during system runtime due to input data or environment changes. Moreover, existing tools cannot recommend applicable configurations to fix the bugs. In comparison, TFix$^+$ adopts the prediction-driven scheme to suggest proper timeout variable configurations.

\section{Conclusion} 
\label{sec:conclusion}

In this paper, we have presented TFix$^+$, a self-configuring hybrid timeout bug fixing system. TFix$^+$ proposes two hybrid schemes to fix misused and missing timeout bugs, respectively. 
TFix$^+$ adopts a prediction-driven timeout configuration recommendation scheme based on runtime tracing and static taint analysis to provide end-to-end bug fixing. 
Our empirical study shows TFix$^+$ can fix 79\% of 91 real world misused or missing timeout bugs. 
The experimental results on 16 reproduced timeout bugs show that TFix$^+$ can produce effective fixes for 15 out of 16 tested timeout bugs. 

\ifCLASSOPTIONcompsoc
  \section*{Acknowledgments}
\else
  \section*{Acknowledgment}
\fi

This research is sponsored in part by NSF CNS1513942 grant, and NSF CNS1149445 grant. 
Any opinions expressed in this paper are those of the authors and do not necessarily reflect the views of NSF.

\small
\bibliographystyle{plain}
\bibliography{icdcs}



\begin{IEEEbiography}[{\includegraphics[width=1in,height=1.25in,clip,keepaspectratio]{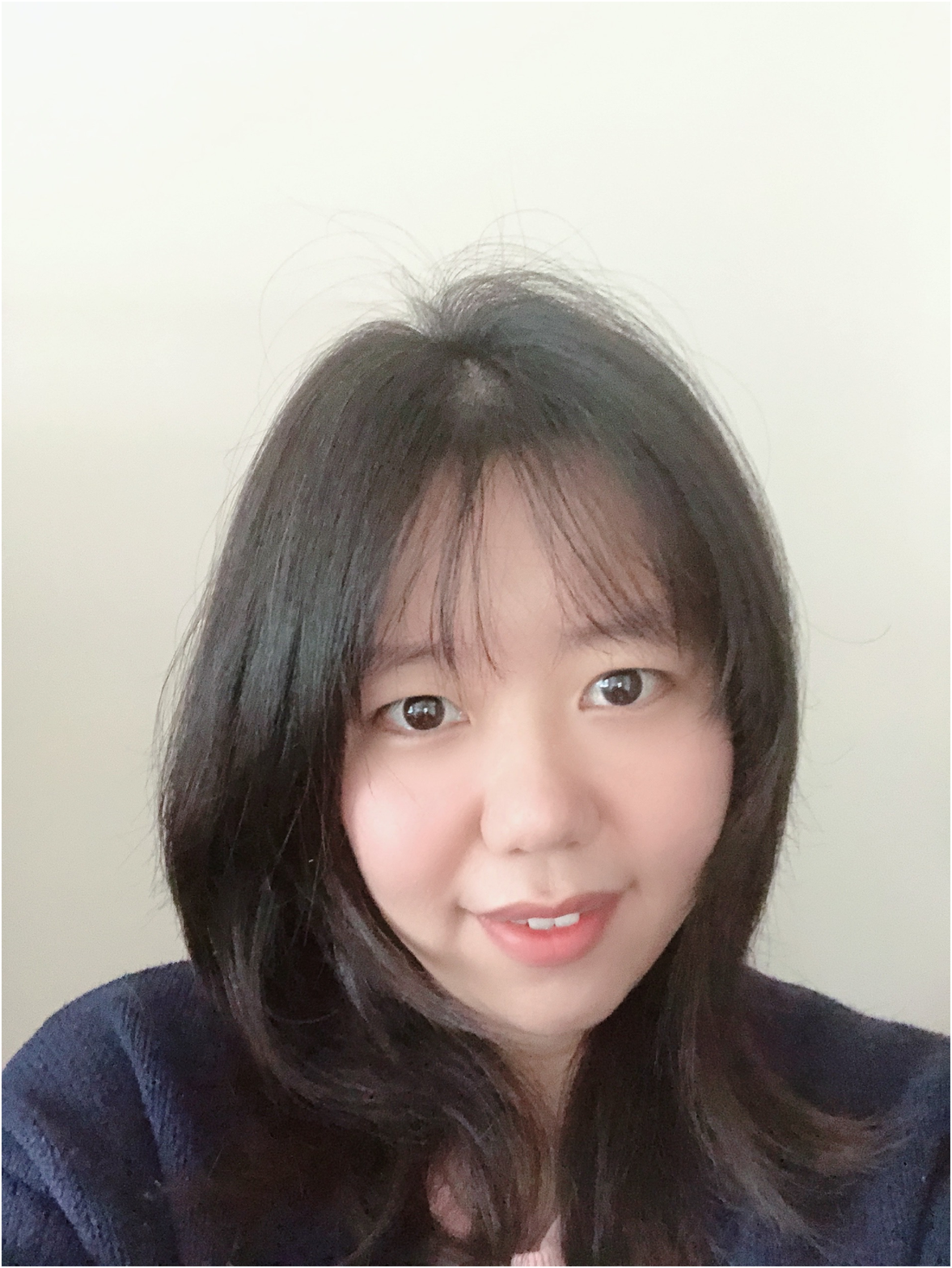}}]{Jingzhu He}
is an assistant professor in the School of Information Science and Technology at the ShanghaiTech University. She received her PhD degree in 2021 from the Department of Computer Science at the North Carolina State University. She received an MPhil degree in computer science from Hong Kong Baptist University in 2016, and her BS degree in electronic and information science from Nanjing University, China in 2013.
\end{IEEEbiography}

\begin{IEEEbiography}[{\includegraphics[width=1in,height=1.25in,clip,keepaspectratio]{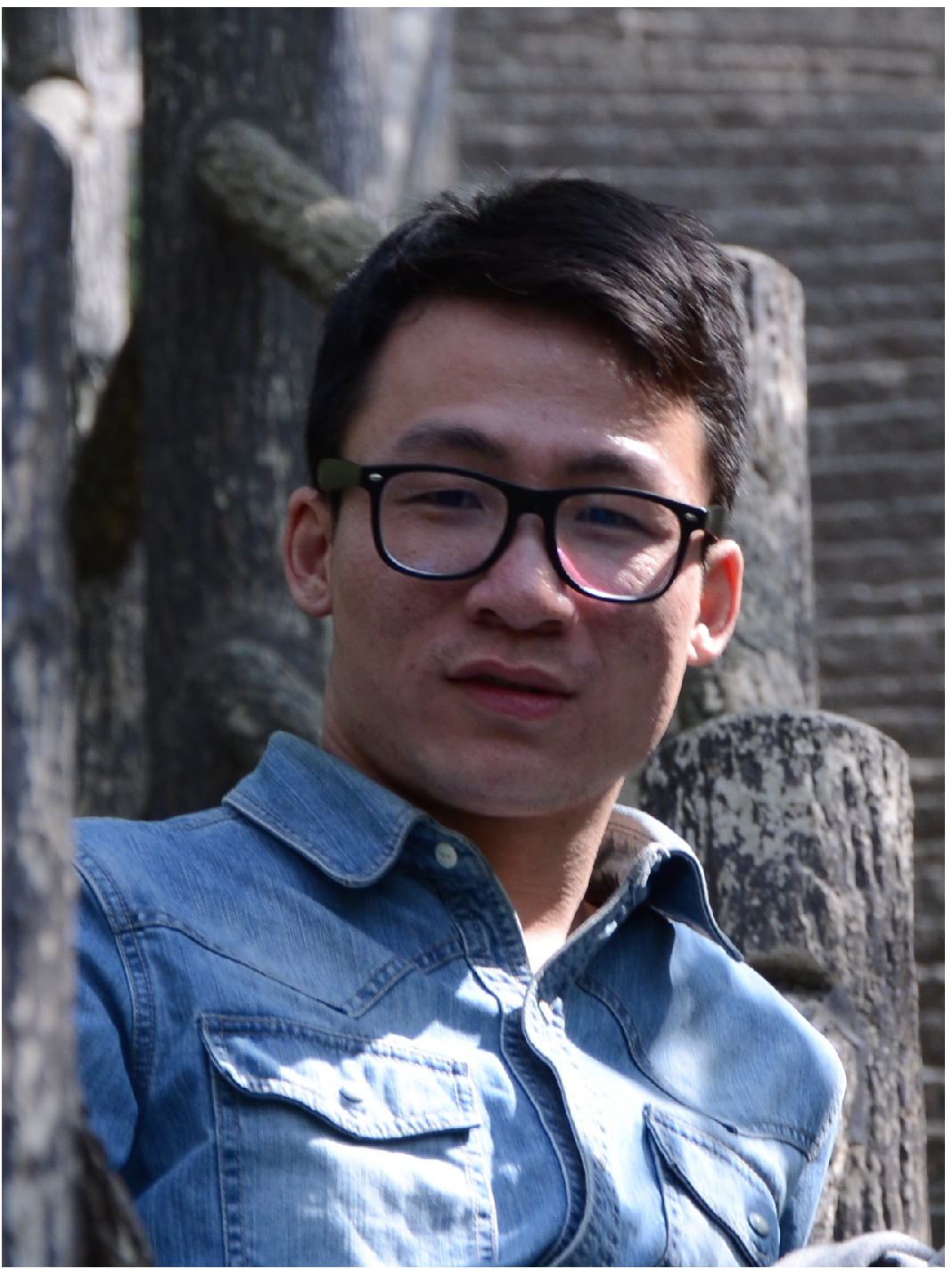}}]{Ting Dai}
is a research staff member at IBM Research. He received his PhD degree in 2019 from the Department of Computer Science at North Carolina State University. He received a BS in information security and MS in computer software and theory from Nanjing University of Posts and Telecommunications, China in 2011 and 2014 respectively. He has interned with IBM Research in the summer of 2018 and InsightFinder Inc. in the summer of 2016. Ting is a member of the IEEE.
\end{IEEEbiography}

\begin{IEEEbiography}[{\includegraphics[width=1in,height=1.25in,clip,keepaspectratio]{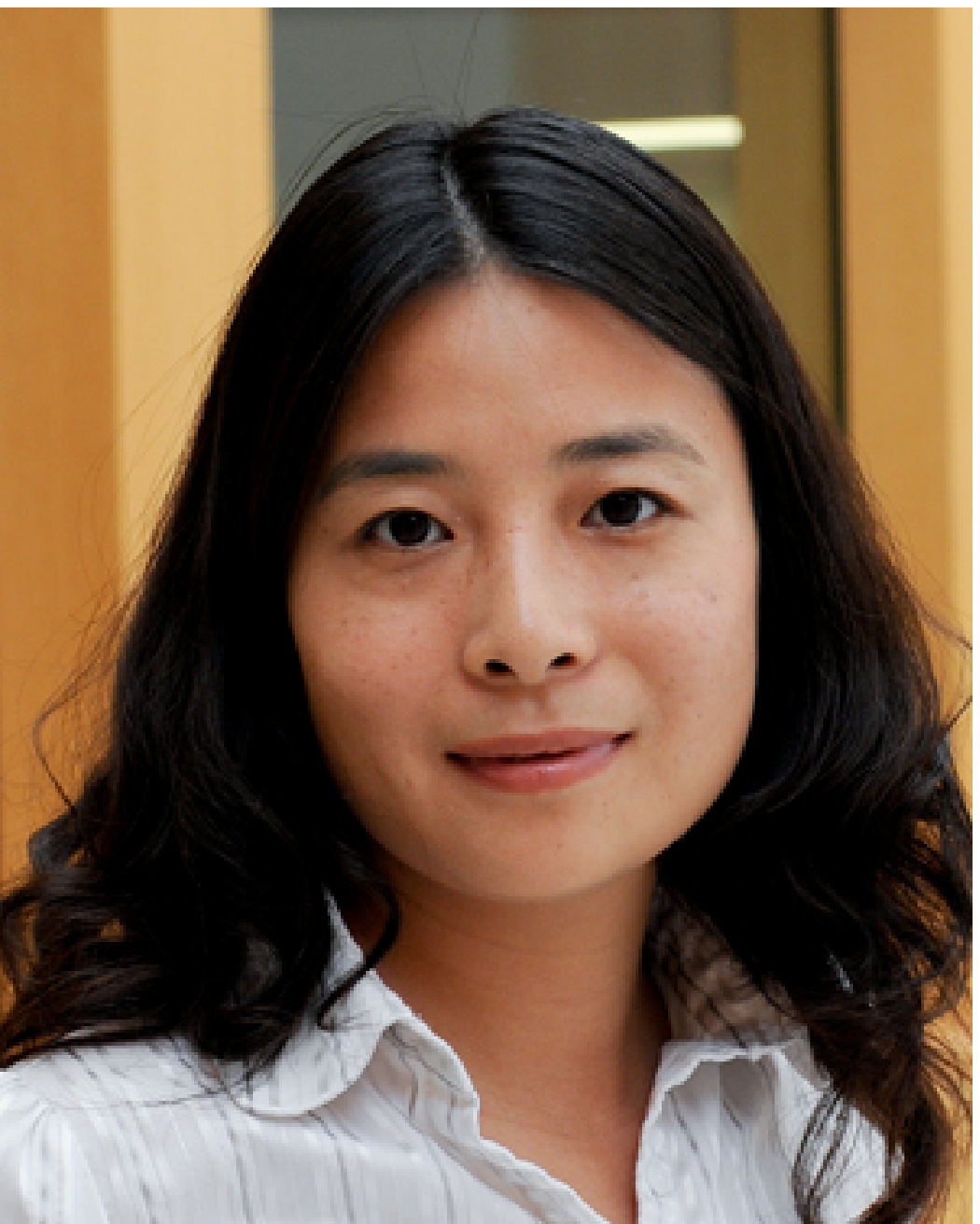}}]{Xiaohui Gu}
is a professor in the Department of Computer Science at the North Carolina State University. She received her PhD degree in 2004 and MS degree in 2001 from the Department of Computer Science, University of Illinois at Urbana-Champaign. She received her BS degree in computer science from Peking University, Beijing, China in 1999. She was a research staff member at IBM T. J. Watson Research Center, Hawthorne, New York, between 2004 and 2007. She received ILLIAC fellowship, David J. Kuck Best Master Thesis Award, and Saburo Muroga Fellowship from University of Illinois at Urbana-Champaign. She also received the IBM Invention Achievement Awards in 2004, 2006, and 2007.  She has filed nine patents, and has published more than 60 research papers in international journals and major peer-reviewed conference proceedings. She is a recipient of NSF Career Award, four IBM Faculty Awards 2008, 2009, 2010, 2011, and two Google Research Awards 2009, 2011, best paper awards from ICDCS 2012 and CNSM 2010, and NCSU Faculty Research and Professional Development Award. She is a Senior Member of IEEE.
\end{IEEEbiography}

\end{document}